\documentclass[aps,prd,10pt,twocolumn,nofootinbib,superscriptaddress]{revtex4-2}
\usepackage{epsfig}
\usepackage{bm}
\usepackage{latexsym}
\usepackage{natbib}
\usepackage{url}
\usepackage[T1]{fontenc}
\usepackage{lmodern}
\usepackage{color}
\usepackage{amsfonts,amssymb,amsmath}
\usepackage{graphicx,epsfig}
\usepackage{psfrag}
\usepackage{subfigure}
\usepackage[citecolor = ProcessBlue]{hyperref}
\hypersetup{colorlinks=true}
\usepackage{mathtools}
\usepackage{enumitem}
\usepackage{float}
\usepackage{mathrsfs}
\usepackage[dvipsnames]{xcolor}
\usepackage[utf8]{inputenc}
\usepackage{comment}
\allowdisplaybreaks
\labelformat{section}{Section~#1} 
\labelformat{equation}{Eq.~(#1)} 
\labelformat{figure}{Fig.~#1} 
\labelformat{subfigure}{Fig.~\thefigure#1} 
\labelformat{table}{Table~#1} 
\begin{document}

\title{Accreting Schwarzschild-like compact object: Plasma-photon interaction and stability}

\author{Avijit Chowdhury}
\email{avijit.chowdhury@rnd.iitg.ac.in}
\affiliation{Department of Physics, Indian Institute of Technology Guwahati,
Guwahati 781039, India}
\affiliation{School of Physical Sciences, Indian Association for the Cultivation of Science, Kolkata 700032, India}

\author{Shauvik Biswas}
\email{shauvikbiswas2014@gmail.com}
\affiliation{School of Physical Sciences, Indian Association for the Cultivation of Science, Kolkata 700032, India}

\author{Sumanta Chakraborty}
\email{tpsc@iacs.res.in}
\affiliation{School of Physical Sciences, Indian Association for the Cultivation of Science, Kolkata 700032, India}
\begin{abstract}

Accretion is a common phenomenon associated with any astrophysical compact object, which is best described by plasma, a state of matter composed of electrons and heavy ions. In this paper, we analyze the linear dynamics of electromagnetic (EM) fields propagating through the accreting plasma around static and spherically symmetric horizon-less, exotic compact objects (ECOs). The general equations governing the propagation of EM waves in such a background exhibits quasi-bound states, whose characteristic frequencies differ from the BH values, for both the axial and the polar modes, as well as for homogeneous and inhomogeneous plasma distributions. Moreover, the real and imaginary parts of these quasi-bound frequencies depict an oscillatory behaviour with the plasma frequency, characteristic of the ECOs considered. The amplitude of these oscillations depend on the non-zero reflectivity of the surface of the compact object, while the oscillation length depends on its compactness. This results into slower decay of the quasi-bound states with time for certain parameter space of the plasma frequency, compared to BHs, making these ECOs more prone to instabilities. 

\end{abstract}
\maketitle
\section{Introduction}

Black holes (BHs) are one of the most remarkable constructs of General Relativity, unmatched in their simplicity and elegance. The characteristic feature of a BH is the existence of an event horizon, acting as a one-way membrane, which causally shields the exterior region from the interior, and hides the singularity within. The observation of gravitational waves (GWs), from the merger of two compact objects, by the LIGO-Virgo-Kagra (LVK) collaboration~\cite{Discovery,PEGW150914,TOGGW150914,GWOSC-3,Cahillane:2022pqm} and the shadow measurement of compact objects at the centre of M87 and our own galaxy, by the Event Horizon Telescope (EHT) collaboration~\cite{Akiyama:2019cqa,EHT-1}, are consistent with these compact objects being BHs. Even though, both of these observations have proved beyond doubt that whatever be the nature of the compact object, it must be more compact than the photon sphere~\cite{Claudel2000JMP}, but the existence of an event horizon has not been directly verified \cite{Cardoso:2016rao,Barack_2019,Cardoso_2019}. Therefore, the question remains whether BHs are the only possible objects which are more compact than a photon sphere\footnote{There can actually be stars, made out of normal matter, which are more compact than the photon sphere, known as the Buchdahl stars \cite{Buchdahl:1959zz, Alho:2022bki, Goswami:2015dma, Chakraborty:2020ifg, Dadhich:2016fku}. However, in the presence of accretion, these stars would describe a significant outflow, which are absent in the EHT observation \cite{EventHorizonTelescope:2022urf, EventHorizonTelescope:2022wok} (however, see \cite{Carballo-Rubio:2022imz}). This suggests that the objects must be more compact than the Buchdahl stars and hence must be made out of exotic matter fields.}, or can there exist other exotic compact objects (ECOs)~\cite{Cardoso:2017njb, Volkel:2017kfj, Oshita:2019sat, Bueno:2017hyj, Cunha:2018gql, Dey:2020lhq}. The exotic-ness of these ECOs stem from the absence of the event horizon in these spacetimes. The ECOs are rather characterised by a surface, close to the event horizon, that is partially/fully reflecting. The reflectivity of the ECO surface, along with its position relative to the BH horizon, encodes information about the object's interior. The ECOs behave exactly as a BH for scales larger than the size of the photon sphere, but the physics at the horizon is completely different. As a consequence there are important effects in the GW sector arising from these ECOs --- (a) existence of echoes in the ringdown waveform of GWs \cite{Maggio:2018ivz,Cardoso:2016oxy,Mark:2017dnq,Tsang:2018uie, Oshita:2018fqu, Konoplya:2018yrp, Dey:2020pth, Mark:2017dnq, Maggio:2019zyv, Biswas:2022wah, Chakraborty:2022zlq,Biswas:2023ofz,Chowdhury:2020rfj}, (b) non-zero tidal Love numbers \cite{Pani:2015tga, Cardoso:2017cfl, Nair:2022xfm, Chakraborty:2023zed}, (c) modifications in the GW phasing due to tidal heating \cite{Datta:2019epe, Chakraborty:2021gdf, Datta:2020rvo}, among many others. Interestingly, most of the above analysis have been performed assuming the ECOs to be isolated, while in reality these objects are in astrophysical environments, be it a dark matter halo, or accreting plasma. The consequences of ECOs evolving in a dark matter environment have been analysed recently \cite{Mitra:2023sny}, while in the present work we focus on the effect of accretion on static and spherically symmetric ECOs. The optical appearances of accretion disks in static and spherically symmetric ECOs has been studied in \cite{Rosa1,Rosa2,Rosa3,Rosa4,Rosa5}.

Just as galactic BHs are supposed to be immersed in a dark matter halo, any astrophysical BHs are also surrounded by accreting matter. Since the accreting matter is essentially made out of plasma, it is expected that electromagnetic (EM) emission from the accreting matter will interact with the plasma, thereby affecting its propagation at large distances. Hence, the most non-trivial behaviour is expected from the EM fields moving in the spacetime. This has already been observed in \cite{cannizzaro2021PRD}, regarding the propagation of EM wave around an accreting Schwarzschild BH, leading to quasi-bound states. This happens because, the photons propagating through the plasma effectively gains a mass, dependent on the plasma frequency $\omega_{\rm pl}$ of the background accreting medium~\cite{SITENKO196719},
\begin{align}\label{eq:plasma-freq}
\omega_{\text{pl}}=\sqrt{\frac{n_\text{e}e^{2}}{m_\text{e}}}~.
\end{align}
Here $n_\text{e}$ is the number density of the electrons, having charge $e$ and mass $m_\text{e}$. In case of inhomogeneous distribution of plasma, the electron number density and the plamsa frequency become dependent on the spacetime coordinates. In the case of static and spherically symmetric background, they become functions of the radial coordinate $r$. The existence of effective mass for photons drastically changes the implications regarding the stability of the compact object~\cite{Rosa2012, pani2012, pani2012-1, Pani2013, Baryakhtar2017, Conlon:2017hhi, Cardoso2018, Frolov2018, Dolan2018, Baumann2019}. It is well-known that non-zero effective mass of the photons may lead to superradiant instabilities for astrophysical BHs~\cite{Brito2015}, and more so for the ECOs, owing to their reflective nature \cite{Maggio:2017ivp}. Therefore, it is important to ask what happens to the photons propagating around an accreting ECO, in particular, whether they lead to any instabilities. As we will demonstrate, the dynamics of the axial part of an EM field is exactly like a Proca field, while the dynamics of the polar sector is very different, which is consistent with the earlier findings in the literature \cite{Cardoso:2020nst, cannizzaro2021PRD}. Another striking difference between the Proca field and EM waves moving in a plasma is that, the motion of EM wave through plasma is still characterised by only two polarization modes (both transverse), whereas a Proca field is always characterised by three polarization modes (one longitudinal and two transverse). Therefore, one must revisit the plasma-photon interaction around ECOs, which have been well understood for BHs~\cite{Cannizzaro:2021zbp, Cannizzaro:2024wnq, cannizzaro2021PRD, PhysRevD.109.023007, Spieksma:2023vwl,Cannizzaro:2024yee}. This is also motivated from the quasi-normal mode (QNM) spectrum for ECOs, where the decay time of the QNMs associated with ECOs exceeded the BH values by several orders of magnitude, making the ECOs prone to instabilities. However, QNMs are eigenvalues of scattering states associated with the compact object, while the EM fields around accreting ECOs lead to quasi-bound states. Therefore, implications for the spectrum of these quasi-bound states associated with an ECO need to be carefully assessed to understand the stability of these objects. As a first step in this direction, we have determined the relevant equations and the frequency spectrum of the quasi-bound states associated with EM waves moving in the background of accreting Schwarzschild-like ECO in this work. We hope to extend our analysis for a rotating ECO and study its stability. 

For this purpose, we start by generalizing the dynamical equations governing the propagation of the EM field through  unmagnetised, cold and collision-less plasma on an arbitrary static and spherically symmetric background. Subsequently, we model the spacetime exterior to the ECO surface by the Schwarzschild metric and investigate how the quasi-bound frequencies respond to the nonzero reflectivity of the ECO surface and its location relative to the horizon. We have considered both constant and frequency-dependent reflectivity, as well as, both the axial and polar part of the EM field. In addition, the case of Bondi accretion, involving inhomogeneous plasma distribution has also been discussed. We have also corroborated our numerical findings with analytic calculations.

The structure of the paper is as follows: we start by defining the general equations governing the motion of an EM wave in a cold and collision-less plasma in \ref{sec:1}. These equations have been separated into axial and polar sectors, and spelled out for a generic static spherically symmetric spacetime, along with the plasma profiles in \ref{sec:2}. Subsequently, in \ref{sec:3} we discuss the numerical setup, methodology and the results obtained. Finally, we conclude with a summary of our results in \ref{sec:4}, highlighting their importance and prospects. Some detailed analytical computations have been delegated to \ref{app:a}, \ref{app-b} and \ref{app-c}, respectively. 

\emph{Notations and conventions:} Throughout the paper, we will use units in which $G=1=c$ and the signature of the flat spacetime Minkowski metric will be taken to be mostly positive, i.e., $\eta_{\mu \nu}=\textrm{diag.}(-1,1,1,1)$. Moreover, all the Greek indices $\mu,\nu,\alpha,\cdots$ denote spacetime indices, and we will exclusively work with four spacetime dimensions. 

\section{Plasma-photon interaction in curved spacetime: Basic equations}\label{sec:1}

In this section we will present the basic equations governing the interaction between plasma and photon in a curved background. We will keep this section brief and the equations general, since they have appeared in various other contexts before \cite{cannizzaro2021PRD,Cannizzaro:2021zbp,Caputo:2021efm,Alonso-Alvarez:2021pgy}, and shall apply them subsequently for a general static and spherically symmetric spacetime. As emphasized before, the interaction between plasma and photon is typical of any accreting system, where the accreting material is best described by a plasma, comprising of electrons and ions, which emits EM waves, interacting with the plasma itself. If within the plasma, the thermal velocity of the electrons is small compared to the propagation speed (phase velocity) of the EM waves, one can classify it as cold plasma. The plasma will be referred to as hot in the reversed situation. Therefore in the cold plasma the following inequality holds true 
\begin{align}
v^{\text{thermal}}_{\rm e}=\sqrt{\frac{2k_{\rm B}T_{\rm e}}{m_{\rm e}}} \ll \frac{\omega}{k}~,
\end{align}
where $\omega$ and $k$ are the frequency and wave number of the EM field, respectively, and $T_{\rm e}$ is the temperature of the electrons within the plasma. In what follows we will concentrate on the case of cold plasma. 

The propagation of EM waves through an unmagnetized cold plasma consisting of electrons and ions (since the ions are much heavier than electrons they barely move compared to the EM waves) in curved spacetime is described by three four vectors --- (a) the vector potential $A_{\mu}$, describing the EM waves, (b) the four velocity $u^{\mu}$ of the electrons, (c) the current density $J^{\mu}$ of the ions, and a scalar --- the number density $n_{\rm e}$ of electrons. The dynamics of the system is thus described by the Maxwell equation together with the momentum and particle conservation equation, \cite{breur1981AA},
\begin{eqnarray}
\nabla_{\nu}F^{\mu\nu}&=&e n_\text{e} u^{\mu} +J^{\mu}~,\label{eq:unpert1}
\\
u^{\mu}\nabla_{\mu}u^{\nu}&=& \frac{e}{m_{\rm e}} F^{\nu}_{\ \mu}u^{\mu}~,\label{eq:unpert2}
\\
\nabla_\mu \big( n_\text{e} u^\mu\big)&=&0~,\label{eq:unpert4}
\end{eqnarray}
where $F_{\mu\nu}=\nabla_{\mu}A_{\nu}-\nabla_{\nu}A_{\mu}$ is the EM field strength tensor and $u^{\mu}u_{\mu}=-1$ with $m_\text{e}$ being the electron mass.

The background spacetime metric will get perturbed by the presence of the accreting plasma, but only at the second order, since the energy-momentum tensor is quadratic in the matter degrees of freedom. For the EM field, this can be easily understood through the following analysis: the vector potential behaves as $A_{\mu}=\epsilon A_{\mu}^{\rm acc}$, where $\epsilon$ is a small parameter and $A_{\mu}^{\rm acc}$ is the vector potential generated by the equilibrium distribution of the accreting plasma. Therefore, the energy momentum tensor of the accreting EM field behaves as $T_{\mu \nu}\sim \epsilon^{2}T_{\mu \nu}^{\rm acc}$, and hence the metric gets perturbed as $g_{\mu \nu}=g_{\mu \nu}^{\rm bg}+\mathcal{O}(\epsilon ^{2})$\footnote{An explicit example would be the case of the Reissner-Nordstr\"{o}m BH, where the vector potential scales as the charge $Q$, while the metric is affected only at $\mathcal{O}(Q^{2})$.}, where $g_{\mu \nu}^{\rm bg}$ is the background static and spherically symmetric metric.   

To assess the stability of the system, we introduce small perturbations over and above the accreting plasma in equilibrium, in particular, we let $A_{\mu}=\epsilon A_{\mu}^{\rm acc}+\epsilon^{2}A_{\mu}^{\rm pert}$. Similar perturbing effects are present to the plasma quantities, namely the electron density and the electron velocity, respectively. Therefore, the dynamical equations for the perturbed EM field $A_{\mu}^{\rm pert}$ can be obtained in terms of $g_{\mu \nu}^{\rm bg}$, $A_{\mu}^{\rm acc}$ and unperturbed quantities associated with plasma in equilibrium. Thus any effect of backreaction on the spacetime metric due to accretion does not affect the dynamics of the photon field $A_{\mu}^{\rm pert}$. As emphasized before, the ions in the plasma are assumed to be heavier than the electrons, and hence any perturbation of the ion quantities is suppressed compared to that of the electron quantities. For this purpose, we project the background metric on the plane orthogonal to the four-velocity of electrons, by introducing $h_{\mu\nu}=g^{\rm bg}_{\mu\nu}+u_{\mu}u_{\nu}$. Thus, any generic tensor can be decomposed into components along and orthogonal to $u_{\mu}$, e.g.~\cite{Ellis:1971pg}, 
\begin{equation}\label{eq:decomp}
\nabla_{\mu} u_{\nu}=\omega_{\mu\nu}+\theta_{\mu\nu}-u_{\mu}u^{\alpha}\nabla_{\alpha}u_{\nu}
\end{equation}
where
\begin{eqnarray}
\omega_{\mu\nu}=\frac{1}{2}\left( v_{\mu\nu} - v_{\nu\mu}\right)\label{eq:omega}~,
\\
\theta_{\mu\nu}=\frac{1}{2}\left( v_{\mu\nu} + v_{\nu\mu}\right)\label{eq:theta}~.
\end{eqnarray}    
Here, $v^{\mu\nu}\equiv h^{\mu\alpha}h^{\nu\beta}\nabla_{\alpha}u_{\beta}$ is the projected four-velocity gradient, with $\omega_{\mu \nu}$ and $\theta_{\mu \nu}$ being the symmetric and the anti-symmetric parts of $v_{\mu \nu}$ (referred to as the vorticity and the deformation tensors, respectively). The electric part of the Maxwell's tensor is given by $E^{\mu}\equiv F^{\mu}_{\ \nu}u^{\nu}$, while in the present context involving un-magnetized cold plasma, the magnetic field identically vanishes. Given \ref{eq:unpert2}, it immediately implies that the acceleration of the electrons are generated by the electric field, as it should. With the above definitions, and following Ref.~\cite{breur1981AA}, we obtain the dynamical equations for the perturbed EM field $A^{\rm pert}_{\mu}$ in an unmagnetized cold plasma, within the Landau gauge ($u^{\mu}A_{\mu}^{\rm pert}=0$) as,
\begin{align}\label{eq:plasma}
&\Big[h^{\alpha \beta}u^{\nu}\nabla_{\nu}\big(\nabla^{\mu}\nabla_{\beta}-\delta^{\mu}_\beta \square \big)
+\big(\omega^{\alpha\beta}+\theta^{\alpha\beta}+\theta h^{\alpha\beta} 
\nonumber
\\
&\qquad +\frac{e}{m_\text{e}}E^\alpha u^\beta \big)\big(\nabla^{\mu}\nabla_{\beta}-\delta^\mu_{\beta}\square\big)
+ \omega_{\text{pl}}^2 h^{\alpha \mu} u^\gamma \nabla_\gamma 
\nonumber
\\
&\qquad +\omega_{\text{pl}}^2 \big(\theta^{\alpha\mu}-\omega^{\alpha\mu}\big)\Big]A_{\mu}^{\rm pert}=0~.
\end{align}
Here, all the raising and lowering of indices have been done by the background metric, e.g., $\theta=g^{\mu \nu}_{\rm bg}\theta_{\mu\nu}$. Note that the Larmour tensor does not appear in the above equation, since we are assuming the plasma to have zero magnetic field. Moreover, \ref{eq:plasma} is in contrast to the effective Proca-like equation, traditionally used to describe the dynamics of photon in plasma, as already pointed out in \cite{cannizzaro2021PRD}. 

\section{EM waves in plasma: Dynamical equations in a generic static and spherically symmetric spacetime}\label{sec:2}

Following the dynamical equations satisfied by an EM field moving in a plasma surrounding an arbitrary curved background, in this section, we apply the same to a generic static and spherically symmetric spacetime. We must emphasize that the spherical symmetry of the background spacetime, and hence of the plasma is a simplifying assumption, which has been used to understand the basic features exhibited by an ECO surrounded by plasma. The results derived here will pave the way for subsequent generalization to the more complex case involving rotating ECOs. The approximation involving static plasma distribution may seem to be in contradiction with the phenomenon of accretion; but is an excellent approximation since the time scales of interest are much shorter than that of accretion. In other words, we assume that the mass of the compact object does not change over the time scale associated with these quasi-bound states. To see that this is indeed a reasonable approximation, we note that for an efficient accretion process the typical time scale corresponds to $t_{\rm accr}\sim (M/\dot{M})\sim 10^{15}\textrm{s}$~\cite{cannizzaro2021PRD,abramowicz2013LRR,Barausse:2014tra}. While the time scale associated with perturbations in the plasma corresponds to $t_{\rm plasma}\sim \omega_{\rm pl}^{-1}\sim 10^{-3}\textrm{s}$, implying $t_{\rm accr}\gg t_{\rm plasma}$. This justifies the use of static configuration for the plasma as well as for the background spacetime.

We take the following line element for the static and spherically symmetric background,
\begin{equation}\label{eq:metric}
ds^2=-f(r)dt^2 +\frac{1}{g(r)}dr^2 +r^2 d\Omega^2~,
\end{equation}
where $d\Omega^2$ denotes the line element on a unit two-sphere. The four velocity of an electron within the plasma in equilibrium is given by $u^{\alpha}=(u^0,0,0,0)$, with $u^0=(1/\sqrt{f})$, since the plasma is assumed to be static, as the time scale of accretion is much longer than the time scale of the quasi-bound states. In this case, the vorticity and the deformation tensor vanish identically. The spherical symmetry of the background spacetime allows us to decompose the vector potential $A_{\mu}^{\rm pert}$ into two independent sectors --- (a) the axial sector, which transforms as $(-1)^{l+1}$ and (b) the polar sector, which transforms as $(-1)^{l}$, under parity. In addition, every component of the vector potential can be expressed as the multiplication of two parts --- one depending on $(t,r)$ coordinates and another on $(\theta,\phi)$ coordinates, respectively. Therefore, after decomposing the vector potential in the spherical harmonic basis, we obtain, 
\begin{align}
A_{\mu}^{\rm pert}&=\sum_{l m}A_{\mu,~l m}^{\rm polar}+A_{\mu,~l m}^{\rm axial}~;
\\
A_{\mu,~l m}^{\rm polar}&=u^{lm}_{1}(t,r)Y_{l m}\delta_{\mu}^{t}+u^{lm}_{2}(t,r)Y_{l m}\delta_{\mu}^{r}
\nonumber
\\
&\qquad +\frac{u^{lm}_{3}(t,r)}{l(l+1)}\left(0,0,\partial_{\theta}Y_{l m},\partial_{\phi}Y_{l m} \right)~,
\label{Apolar}
\\
A_{\mu,~l m}^{\rm axial}&=\frac{u^{lm}_{4}(t,r)}{l(l+1)}\left(0,0,\frac{1}{\sin \theta}\partial_{\phi}Y_{l m},-\sin \theta \partial_{\theta}Y_{l m}\right)~.
\label{Aaxial}
\end{align}
The above angular functions ensure that under parity transformation the functions $u^{lm}_{i}$ belong to the polar sector for $i=1,2,3$, and to the axial sector for $i=4$. Thus, the EM field has been neatly separated into axial and polar sectors.

\subsection{Axial Sector}\label{sec:2a}

In the axial sector, we have a single equation for $u^{lm}_{4}$ to solve. This equation can be obtained by substituting $A_{\mu}^{\rm axial}$ from \ref{Aaxial} to \ref{eq:plasma}, resulting into the following dynamical equation for $u^{lm}_{4}(t,r)$,
\begin{align}\label{eq:pl-u4}
&f(r)\Big[-2l(l+1)+r^2 \omega_{\text{pl}}^{2}\Big]\frac{\partial u_{4}^{lm}(t, r)}{\partial t} 
\nonumber
\\
&\quad +r^2\bigg[g'\frac{\partial^{2}u_{4}^{lm}(t, r)}{\partial t\partial r}
+2g\frac{\partial^{3}u_{4}^{lm}(t, r)}{\partial t\partial r^{2}}\bigg]
\nonumber
\\
&\quad +r^2\left[gf'\frac{\partial^{2}u_{4}^{lm}(t, r)}{\partial t\partial r}  
-2\frac{\partial^{3}u_{4}^{lm}(t, r)}{\partial t^{3}}\right]=0~.
\end{align}
Here, `prime' denotes derivative with respect to $r$. Note that, the plasma frequency $\omega_{\rm pl}$ depends on the radial coordinate $r$ through the number density of electrons. Defining, $\partial_t u^{lm}_{4}\equiv \Psi^{lm}_{\text{ax}}(t,r_*)$, we can rewrite the \ref{eq:pl-u4} as,
\begin{equation}\label{eq:pl-odd-time}
\frac{d^2\Psi^{lm}_{\text{ax}}}{dr_{*}^{2}}-\frac{d^2\Psi^{lm}_{\text{ax}}}{dt ^2}-\frac{f(r)}{r^2}\Big[l(l+1)+r^2 \omega_{\text{pl}}^2\Big]\Psi^{lm}_{\text{ax}}=0~,
\end{equation}
where $r_*$ is the tortoise coordinate, defined as
\begin{equation}\label{eq:trts}
\frac{dr_*}{dr}=\frac{1}{\sqrt{f(r)g(r)}}~.
\end{equation}
The above equation can be expressed as a Schr\"{o}dinger-like equation by transforming to the frequency domain. This is achieved by the expressing the axial master variable as, $\Psi^{lm}_{\rm ax}(t,r)=\int d\omega e^{-i\omega t}\Psi^{lm}_{\rm ax}(\omega,r)$, such that we obtain the following equation,
\begin{equation}\label{eq:pl-odd}
\frac{d^2 \Psi^{lm}_{\text{ax}}}{dr_{*}^{2}}+\left(\omega^2-V^{lm}_{\text{ax}}\right)\Psi_{\text{ax}}=0~.
\end{equation}
where, the effective potential takes the following form,
\begin{equation}\label{eq:V-pl-odd}
V^{lm}_{\text{ax}}=\frac{f(r)}{r^2}\Big[l(l+1)+r^2 \omega_{\text{pl}}^2\Big]~.
\end{equation}
Intriguingly, the axial potential does not depend on $m$, and in the limit $\omega_{\rm pl}\to 0$, we get back the effective potential experienced by a photon in a static and spherically symmetric background. Even though the differential equation satisfied by $A_{\mu}^{\rm pert}$ is complicated, at least for the axial sector, it resembles the effective potential of a Proca field with mass $m\sim \omega_{\rm pl}$~\cite{Rosa2012,cannizzaro2021PRD}.

\subsection{Polar Sector}\label{sec:2b}

In the polar sector, we have three unknown functions $u^{lm}_{1}$, $u^{lm}_{2}$ and $u^{lm}_{3}$, whose evolution equations are to be determined from the general differential equation for $A_{\mu}^{\rm pert}$. Since we have chosen the four velocity of electrons in equilibrium to have only the zeroth component, the Landau gauge condition ensures that, 
\begin{equation}\label{eq:u1}
u^{lm}_1 =0~.
\end{equation}
Therefore, instead of three we now have two EM degrees of freedom. In order to determine the differential equations associated with the evolution of $u^{lm}_{2}$ and $u^{lm}_{3}$, we substitute \ref{Apolar} in \ref{eq:plasma}, which along with the condition presented in \ref{eq:u1}, yields the following two coupled equations,
\begin{align}
\Big[l(l+1)&+r^2\omega_{\text{pl}}^2\Big]\frac{\partial u_{2}^{lm}}{\partial t} 
-\frac{\partial^{2}u_{3}^{lm}}{\partial t\partial r}
+\frac{r^2}{f(r)}\frac{\partial^{3}u_{2}^{lm}}{\partial t^{3}}=0~,
\label{eq:u2}
\\
f(r)\Bigg[2\omega_{\text{pl}}^2&\frac{\partial u_{3}^{lm}}{\partial t} 
+g'\Big\{l(l+1)\frac{\partial u_{2}^{lm}}{\partial t}
-\frac{\partial^{2}u_{3}^{lm}}{\partial t\partial r}\Big\}\Bigg]
\nonumber
\\
+g\Bigg[f'\Big\{&l(l+1)\frac{\partial u_{2}^{lm}}{\partial t}
-\frac{\partial^{2}u_{3}^{lm}}{\partial t\partial r}\Big\} 
+2f\Big\{-\frac{\partial^{3} u_{3}^{lm}}{\partial t\partial r^{2}}
\nonumber
\\
&+l(l+1)\frac{\partial^{2} u_{2}^{lm}}{\partial t\partial r}\Big\}\Bigg] 
+ 2 \frac{\partial^{3} u_{3}^{lm}}{\partial t^{3}}=0~.
\label{eq:u3}
\end{align}
Note that the above equations are third order differential equations, which is consistent with the order of the differential equation for $A_{\mu}^{\rm pert}$, as presented in \ref{eq:plasma}. To proceed further, and to obtain a second order differential equation, we switch to the frequency domain, i.e., we express the two variables $u_{2}^{lm}$ and $u_{3}^{lm}$ as, 
\begin{align}
u_2^{lm}(t,r)&=\int d\omega U_2^{lm}(\omega,r) e^{-i\omega t}~,
\\
u_3^{lm}(t,r)&=\int d\omega U_3^{lm}(\omega,r)e^{-i\omega t}~.
\end{align}
Therefore, \ref{eq:u2} can be used to express $U_{2}^{lm}$ in terms of the radial derivative of $U_{3}^{lm}$ as,
\begin{equation} \label{eq:U2}
U^{lm}_{2}=\frac{f{U^{lm}_{3}}'}{-r^{2}\omega^{2}+f\{l(l+1)+r^{2}\omega_{\text{pl}}^{2}\}}~.
\end{equation}
Since in the limit, $f(r)\to 0$, the EM degree of freedom $U_{2}^{lm}$ identically vanishes, we should find out the master differential equation in terms of $U_{3}^{lm}$. Substituting the above expression for $U^{lm}_2$ in the frequency domain representation of \ref{eq:u3}, the following decoupled equation for $U_{3}^{lm}$ can be obtained,
\begin{widetext}
\begin{align}\label{eq:A1-inhomo}
{U_{3}^{lm}}''&+\frac{1}{2rfg\Bigl(-\omega^2+f\omega_{\text{pl}}^2\Bigr)
\biggl[-r^2\omega^2+f\Bigl\{l(l+1)+r^2 \omega_{\text{pl}}^2\Bigr\}\biggr]}
\Bigg[r^3 \omega^4 gf'+f\biggl(r\omega^2 g\Bigl\{l(l+1)-2r^2\omega_{\text{pl}}^2\Bigr\}f'+r^3 \omega^4 g'\biggr)
\nonumber
\\ 
&+f^2\biggl(g\Bigl\{-4l(l+1)\omega^2+l(l+1)rf'\omega_{\text{pl}}^2+r^3f'\omega_{\text{pl}}^4\Bigr\} 
-r\omega^2 \Bigl\{l(l+1)+2r^2\omega_{\text{pl}}^2\Bigr\}g'\biggr)
+f^3\omega_{\text{pl}}\biggl(r\omega_{\text{pl}}\Bigl\{l(l+1)+r^2\omega_{\text{pl}}^2\Bigr\}g'
\nonumber
\\ 
&+4l(l+1)g\bigl\{\omega_{\text{pl}}+r\omega_{\text{pl}}'\bigr\}\biggr)\bigg]{U_{3}^{lm}}'
+\frac{r^2\omega^2-f\left\{l(l+1)+ r^2 \omega_{\text{pl}}^2\right\}}{r^2gf}U_{3}^{lm}=0~.
\end{align}
\end{widetext}
Note that the above equation involves derivatives of the plasma frequency, since we have assumed the plasma properties to be a function of the radial coordinate. In the case of homogeneous plasma, satisfying $\omega_{\text{pl}}=\textrm{constant}$, in a Schwarzschild background, with $g=f=1-(2M/r)$, one can easily verify that \ref{eq:A1-inhomo} reduces to Eq.~(A1) of Ref.~\cite{cannizzaro2021PRD}. In the general case as well, i.e., with $f\neq g$, it is possible to recast \ref{eq:A1-inhomo} in a Schrodinger-like form, which reads
\begin{equation}\label{eq:pl-even}
\frac{d^2\Psi^{lm}_{\text{pol}}}{dr_{*}^{2}}-V^{lm}_{\text{pol}}\Psi^{lm}_{\text{pol}}=0~,
\end{equation}
where, $r_{*}$ is the Tortoise coordinate, defined in \ref{eq:trts}. In order to express the polar sector differential equation to the above form, one must rescale $U_{3}^{lm}$, so that the first derivative term in \ref{eq:A1-inhomo} goes away. This is achieved by introducing the master function $\Psi^{lm}_{\text{pol}}$, for the polar sector, which is defined as, $\Psi_{\text{pol}}\equiv G(r)U_3^{lm}$ with\footnote{The function $G(r)$ is defined as the inverse of the auxiliary function defined in \cite{cannizzaro2021PRD}; the homogeneous plasma profile can be obtained from the above equation by considering $\omega_{\text{pl}}$ to be a constant.}
\begin{equation}\label{eq:aux-polar}
G(r)=r\sqrt{\frac{f(r)\omega_{\text{pl}}^2-\omega^2}{f(r)\big\{l(l+1)+r^2\omega_{\text{pl}}^2\big\}-r^2\omega^2}}~.
\end{equation}
Therefore, alike the odd sector, in the even sector as well, we have only one propagating degree of freedom. Therefore, as in the massless case, here also we have two propagating degrees of freedom. This is in striking contrast with the Proca case, where one has three propagating degrees of freedom. The complete expression for the effective potential $V_{\rm pol}$ in the polar sector can be found in \ref{app:a}. Therefore, \ref{eq:pl-odd} and \ref{eq:pl-even}, together with \ref{eq:V-pl-odd} and \ref{eq:V-pl-even} in \ref{app:a} govern the photon dynamics through cold and unmagnetised plasma distribution in the background of a generic static and spherically symmetric spacetime geometry.
\subsection{Plasma Profile}\label{sec:2c}

After describing the evolution equations for the axial and polar parts of EM waves, we present here the details of the plasma profile considered in this work. We will work with two distinct plasma profiles:
\begin{itemize}

\item A homogeneous density profile, characterised by a constant electron/ion density, yielding a constant plasma frequency $\omega_{\rm pl}$. Therefore, the derivative $\omega_{\rm pl}'$, along with all the higher derivatives identically vanishes.  

\item An inhomogeneous distribution of plasma, dependent on the radial coordinate $r$, which may arise in a Bondi-like accretion disk model, describing the accretion dynamics of a non-self-interacting gas, with density $\rho$, and pressure $p$, around a spherically symmetric compact object. 
In this context, the motion of plasma particles are considered to be spherically symmetric, steady and adiabatic. 
However, these particles generically have non-zero radial velocity,
which can turn supersonic below the radius $r_{\rm s}$, satisfying: $2v_{\rm s}^{2}r_{\rm s}=M$, where, $v_{\rm s}=\sqrt{(\gamma p/\rho)}$, with $\gamma$ being the polytropic index of the gas~\cite{Choudhuri_1998,Ray:2006ydw}. 
Typically, this radius $r_{\rm s}$, known as the sonic point, appears near the compact object, and hence there will be substantial radial velocity near the ECO surface. Since the observations are performed over a time scale $t_{\text{plasma}}\ll t_{\text{accr}}$, and hence the number density of plasma particles radially in-falling towards the ECO is very small (in other words, the rate of accretion $\dot{M}$ is negligible), as a first approximation we will ignore these radial velocities, and shall assume the four velocity of the plasma particles to be given by $u^{\mu}=(1/\sqrt{f},0,0,0)$.

The Bondi-like accretion model predicts a  power-law density profile of the electron/ion and consequently a plasma frequency of the form,
\begin{equation}\label{eq:bondi}
\omega^2_{\text{pl}}(r)=\omega^2_B\left(\frac{r_0}{r}\right)^\lambda+\omega^2_{\infty}~,
\end{equation}
where, $r_{0}$ is the radius of the compact object, such that $\sqrt{\omega^2_B +\omega^2_{\infty}}$ is the plasma frequency at the surface of the compact object. For example, in the case of a BH as well as exotic-compact object, one may consider $r_0$ to be replaced by $r_{+}$ in \ref{eq:bondi}. Moreover, $\omega_{\infty}$ is the plasma frequency at infinity, such that $\omega_B \ll \omega_{\infty}$. Finally, the parameter $\lambda$ depends on the adiabatic index of the gas, e.g., $\lambda=(3/2)$ for a monoatomic gas.

\end{itemize}

\subsection{Nature of the central compact object}\label{sec:2d}

In this section, we will describe the nature of the central compact object around which the plasma is accreting, and whose exterior geometry is described by the static and spherically symmetric line element as in \ref{eq:metric}. The analysis assuming the compact object to be a BH has been performed in \cite{cannizzaro2021PRD}, while here we focus on the case of ECO. These objects are ultra compact and have radii close to the BH horizon. The difference between the surface radius of an ECO from that of a BH is one of the characteristic properties of the ECO and is traditionally denoted by $\epsilon$, with the following definition:
\begin{equation}
\epsilon \equiv \frac{r_{0}-r_{+}}{r_{+}}~.
\end{equation}
Here, $r_{0}$ is the surface radius of the ECO and $r_{+}$ denotes the location of the would be horizon\footnote{This is obtained by solving the equation $f(r)=0=g(r)$. If the zeroes of $f(r)$ and $g(r)$ do not coincide, then rather than a BH it will denote a wormhole spacetime, another class of ECO.}. We will consider cases with $\epsilon \ll 1$, and this requires either some exotic matter field or, some quantum effects at play, since normal matter fields cannot give rise to such compact structures. Furthermore, since the horizon is replaced by a surface, it is expected that the EM waves will not be fully absorbed, as it would have been for a BH, rather will also have a reflected component. The reflectivity denotes another characteristic feature of an ECO. 

To quantify the reflectivity, we note that the effective potential $V_{\rm ax}^{lm}$ in the axial sector, presented in \ref{eq:V-pl-odd}, vanishes in the limit $f(r)\to 0$, i.e., $r\to r_{+}$. Therefore, \ref{eq:pl-odd} has plane waves $e^{\pm i\omega r_{*}}$ as solutions. The same holds true for the polar sector as well, where $V_{\rm pol}^{lm}$ will go to $-\omega^{2}$ in the limit $r\to r_{+}$. Therefore, \ref{eq:pl-even} will also have plane wave modes as solutions. For a BH, only ingoing waves will exist near the horizon, and hence the solution will be given by, $e^{-i\omega r_{*}}$. While for an ECO there will be a reflected part and hence the general solution will be, 
\begin{equation}
\Psi_{\rm ax/pol}^{lm}=Ae^{-i\omega r_{*}}+Be^{i\omega r_{*}}~.
\end{equation}
Following which, one may define the reflectivity of an ECO in the present context of static and spherically symmetric background geometry as, 
\begin{equation}\label{def_reflect}
R(\omega)\equiv \left[\frac{1-\dfrac{i}{\omega}\left(\dfrac{d\ln \Psi^{lm}_{\rm ax/pol}}{dr_{*}}\right)}{1+\dfrac{i}{\omega}\left(\dfrac{d\ln \Psi^{lm}_{\rm ax/pol}}{dr_{*}}\right)}\right]_{r_{0}}
=\frac{B}{A}e^{2i\omega r^{0}_{*}}~,
\end{equation}
where, $r^{0}_{*}=r_{*}(r_{0})$. Therefore, the ratio of the outgoing and ingoing amplitudes of EM waves near the surface of the ECO is proportional to the reflectivity, modulo an overall phase factor. Therefore any model of an ECO will be characterized by the two parameter family $(\epsilon, R)$. Note that, in general, the reflectivity is a function of the frequency $\omega$ of EM waves. For example, in the case of BHs with quantum corrections, it acquires a reflectivity: $R=\exp(-\hbar \omega/k_{\rm B}T_{\rm H})$, where $T_{\rm H}$ is the Hawking temperature. On the other hand, in the context of the membrane paradigm, the dominant contribution to the reflectivity is given by, $|R|=(1-16\pi \eta)/(1+16\pi \eta)$. In the present context as well, we will present our results for constant reflectivity, as well as for the Boltzman reflectivity. 

\section{Plasma-driven quasi-bound modes for Schwarzschild-like ECO}\label{sec:3}

Having described all the necessary ingredients, we now present the analysis involving the stability of quasi-bound states for the EM waves propagating within a plasma, which is acccreting around an ECO. For the QNMs, which are scattering states, it is well-known that ECOs lead to a very small value for the imaginary part of QNMs, responsible for the echoes in the GW ringdown. On the other hand, for the quasi-bound states, the difference between BHs and ECOs have not been studied extensively (however, see \cite{Cardoso_2019,PhysRevD.108.103025}). In particular, if any compact object is accreting, which is common for most of the compact objects in an astrophysical environment, such quasi-bound states will naturally form. Thus it is important to understand these quasi-bound states and their implication for stability, if the central compact object is not a BH, but an ECO.  

To model the spherically symmetric ECO, we assume that the spacetime outside the surface of the ECO is described by a Schwarzschild metric, with the metric elements
\begin{equation}\label{eq:sch-eco}
f(r)=g(r)=1-\frac{2M}{r}~,
\end{equation}
where $r\geq r_{0}$, with $r_0=2M(1+\epsilon)$ being the location of the surface of the ECO. Note that our previous analysis involving the evolution equations for the EM field in the axial and the polar sector are completely general and can be applied to any generic static and spherically symmetric background. However, to see the basic features of the ECO on the quasi-bound states we have kept the background spacetime simplest, namely Schwarzschild. It is possible to generalize our analysis to any other static and spherically symmetric geometry, including wormholes. 

\subsection{Numerical Method}\label{sec:3a}

In this section we provide a brief analysis of the numerical method we have adopted for determining the characteristic frequencies of plasma-driven quasi-bound states for EM field in the background of an ECO. Throughout this paper, we use the direct integration (shooting) method~\cite{Rosa2012}. The idea is to integrate the system of differential equations forward from the surface of the compact object and backwards from a large radius (treated as infinity) to a common matching point. The matching between the two solutions is achieved by demanding that the Wronskian of the two solutions vanish at the matching point. This yields the frequencies of the quasi-bound modes. The procedure requires suitable boundary conditions to be specified close to the surface of the compact object and near asymptotic infinity,
however, it is not specific to the nature of the effective potential and thus has versatile applicability.

Near the surface of the compact object, which has both ingoing and outgoing modes present in the analysis, the solution of the differential equations governing the axial and polar modes, may be written as the sum of a purely ingoing wave and a purely outgoing wave, 
\begin{align}\label{eq:gen-bc-wall}
\Psi^{lm}_{\text{ax}/\text{pol}}&\sim e^{-i\omega(r_*-r^{0}_*)}\sum_{n}h^{\text{ax}/\text{pol}}_n (r-2M)^n
\nonumber
\\
&+R(\omega) e^{i \omega (r_{*}-r^{0}_{*})} \sum_{n}k^{\text{ax}/\text{pol}}_n (r-2M)^n~.
\end{align}
Here, $R(\omega)$ is the reflectivity of the surface of the ECO, defined in \ref{def_reflect}. Note that $r=2M$ is a regular singular point of the evolution equations for the axial, \ref{eq:pl-odd}, and the polar, \ref{eq:pl-even}, sectors. Hence the series around $r=2M$ has a radius of convergence $\gg\epsilon$, so that the above series applies to the ECO as well.

Just as the above series describes the master functions for polar and axial sectors near the surface of the ECO, near asymptotic infinity as well, similar power series expansion of the solutions are possible and they are generically given by,
\begin{equation}\label{eq:gen-bc-infty}
\Psi^{lm}_{\rm ax/pol} \sim \sum_{n} \frac{B_n^{\text{ax}/\text{pol}}}{r^n} e^{-k_\infty r_*}+ \sum_{n} \frac{C_n^{\text{ax}/\text{pol}}}{r^n} e^{k_\infty r_*}~,
\end{equation}
where $k_{\infty}= \sqrt{\omega_{\infty}^2-\omega^2}$, with $\omega_{\infty}$ being the plasma frequency at asymptotic infinity. Since we are interested in bound states, which are strongly localized near the ECO, their amplitudes at large distance must be exponentially small. Therefore, for the existence of quasi-bound states we assume that --- (a) $\omega< \omega_{\infty}$, and (b) $C_n^{\text{ax}/\text{pol}}=0$. This corresponds to the exponentially decaying solution near the spatial infinity, which are highly localized near the compact object.

Using the approach and the boundary conditions specified above, in the following section, we will discuss the fundamental quasi-bound frequencies of the EM waves propagating on the background of a Schwarzschild-like ECO in the presence of accreting plasma. We would like to point out that tracking the fundamental mode (along with overtones) using the direct integration method becomes extremely challenging for small $M\omega_{\text{pl}}$, as the imaginary part of the quasi-bound frequency becomes very small, and hence prone to numerical errors. This applies to both BHs and ECOs. However, in the case of ECOs, there are additional numerical challenges, to be described later. The numerical accuracy of the results in general is also dependent on the truncation order of the series at both the boundaries. To ensure the robustness of our numerical results: (a) we verified that the frequencies obtained are independent of the location of the matching point, (b) for each case, we ensured that the change induced in the frequencies due to (reasonable) change in the truncation order is below the precision of the obtained results and (c) in each case, we scanned the parameter space in small steps to ensure that same overtone mode is tracked through the parameter space.

\subsection{Quasi-bound states: Axial sector}\label{sec:3b}

\begin{figure}[h]
\includegraphics[width=\linewidth]{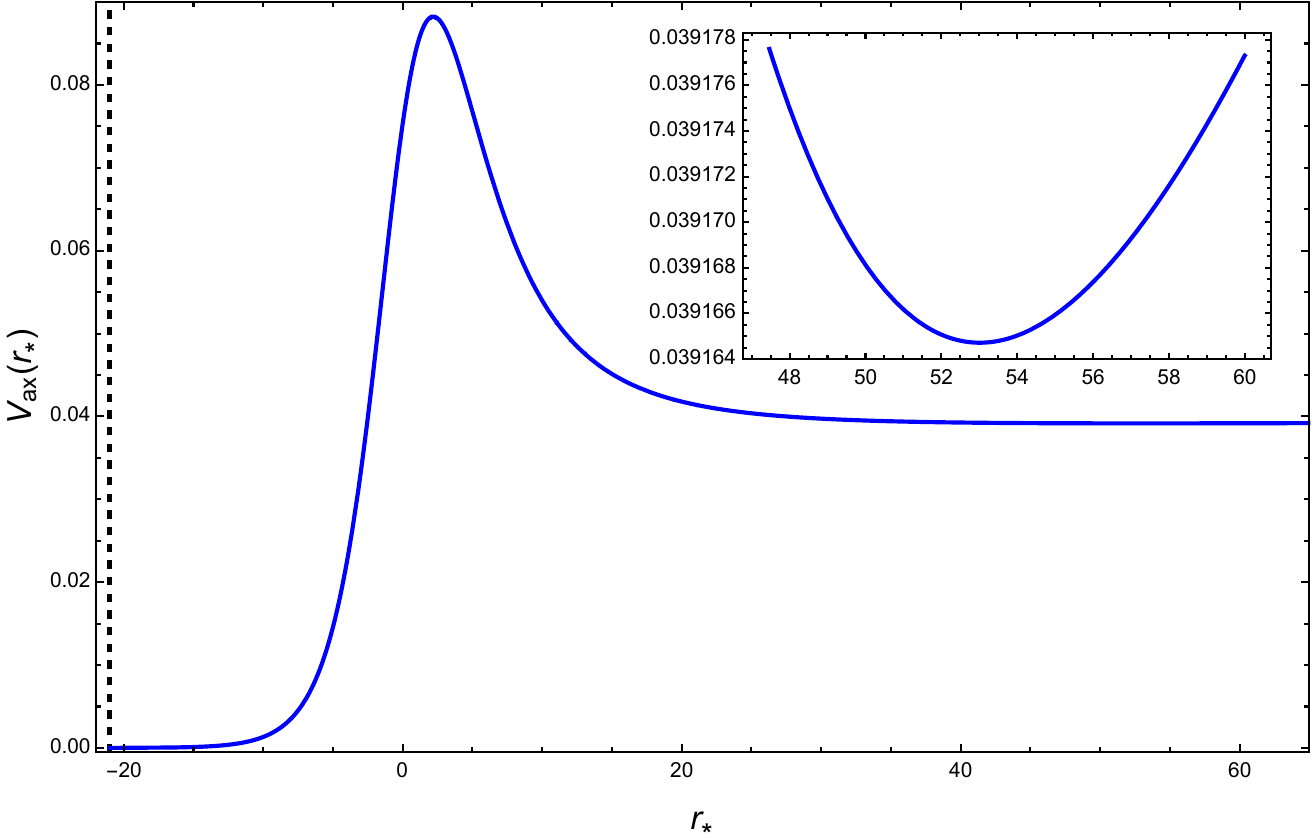}
\caption{Plot of the effective potential $V_{\rm ax}$ in the axial sector for a Schwarzschild-like ECO (with $M=1$) as a function of the Tortoise coordinate $r_*$, with $M \omega_{\text{pl}}=0.2$, $\epsilon=10^{-5}$ and $l=1$. The dashed line represents the surface of the ECO, while the inset figure shows the minima of the effective potential resulting in bound states. Note that the potential vanishes near the surface of the ECO, but approaches to a constant value $\omega_{\rm pl}^{2}=0.04$ asymptotically.}\label{fig:pot}
\end{figure}

With the numerical scheme and appropriate boundary conditions at hand, let us determine the quasi-bound states associated with the axial sector of EM waves. We start by assuming a homogeneous plasma profile, whose effective potential for a Schwarzschild-like ECO with mass $M$ is given by
\begin{equation}\label{eq:ax-sch}
V^{lm}_{\text{ax}}=\left(1-\frac{2M}{r}\right)\left(\omega_{\text{pl}}^{2}+\frac{l(l+1)}{r^{2}}\right)~.
\end{equation}
For quasi-bound state to exist for both BHs and ECOs, the above potential must have two extrema --- one maxima and one minima, and the maxima must be closer to the compact object than the minima (see \ref{fig:pot} for an example). Taking derivatives of the effective axial potential $V^{lm}_{\text{ax}}$, it follows that, the above conditions are satisfied for $M\omega_{\rm pl}<\sqrt{l(l+1)/12}$ (for a derivation, see \ref{app-b}). Thus, there exists an upper bound for the plasma frequency. 

\begin{figure*}[htbp!]
\includegraphics[width=0.45\textwidth]{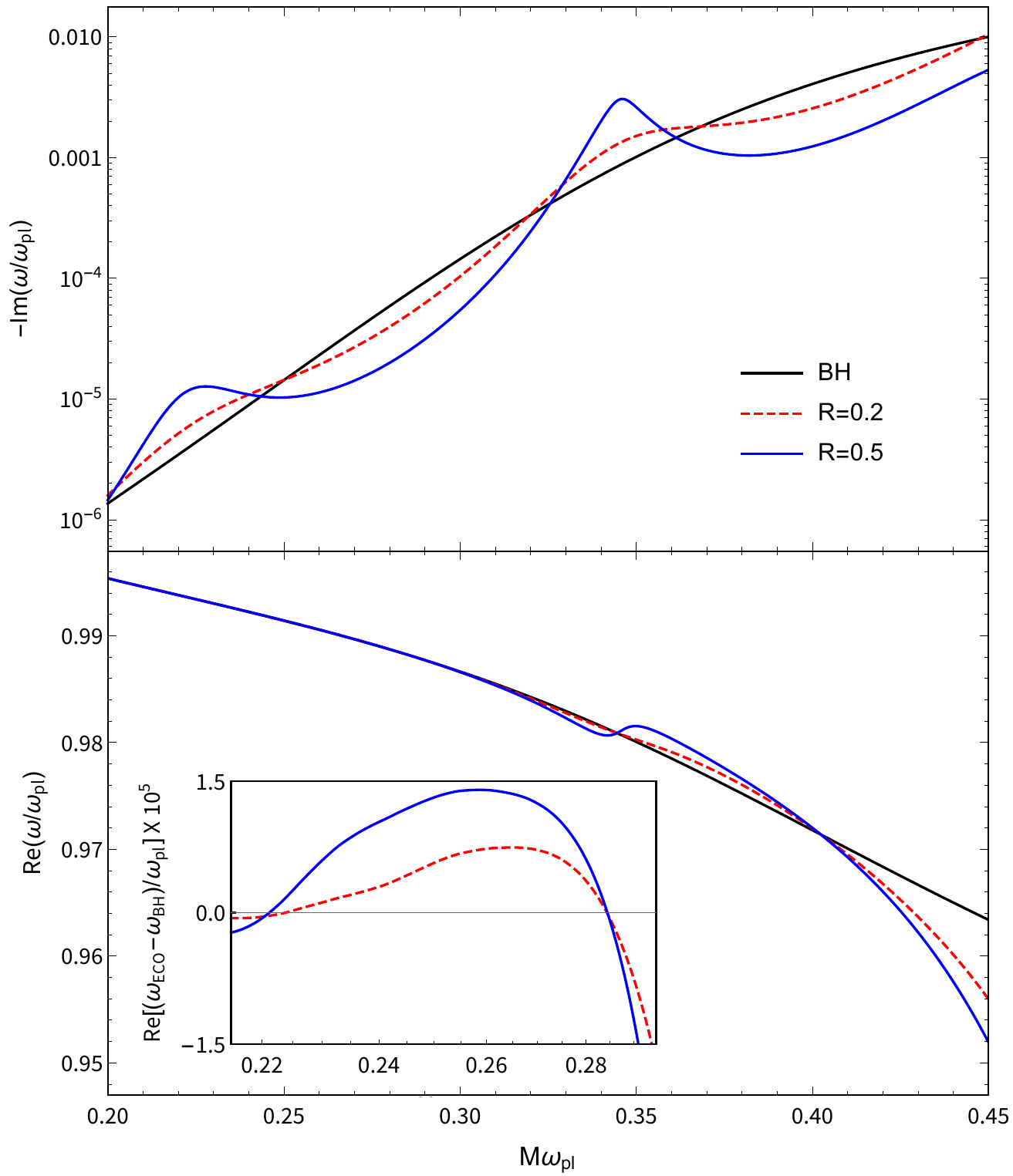}
\includegraphics[width=0.45\textwidth]{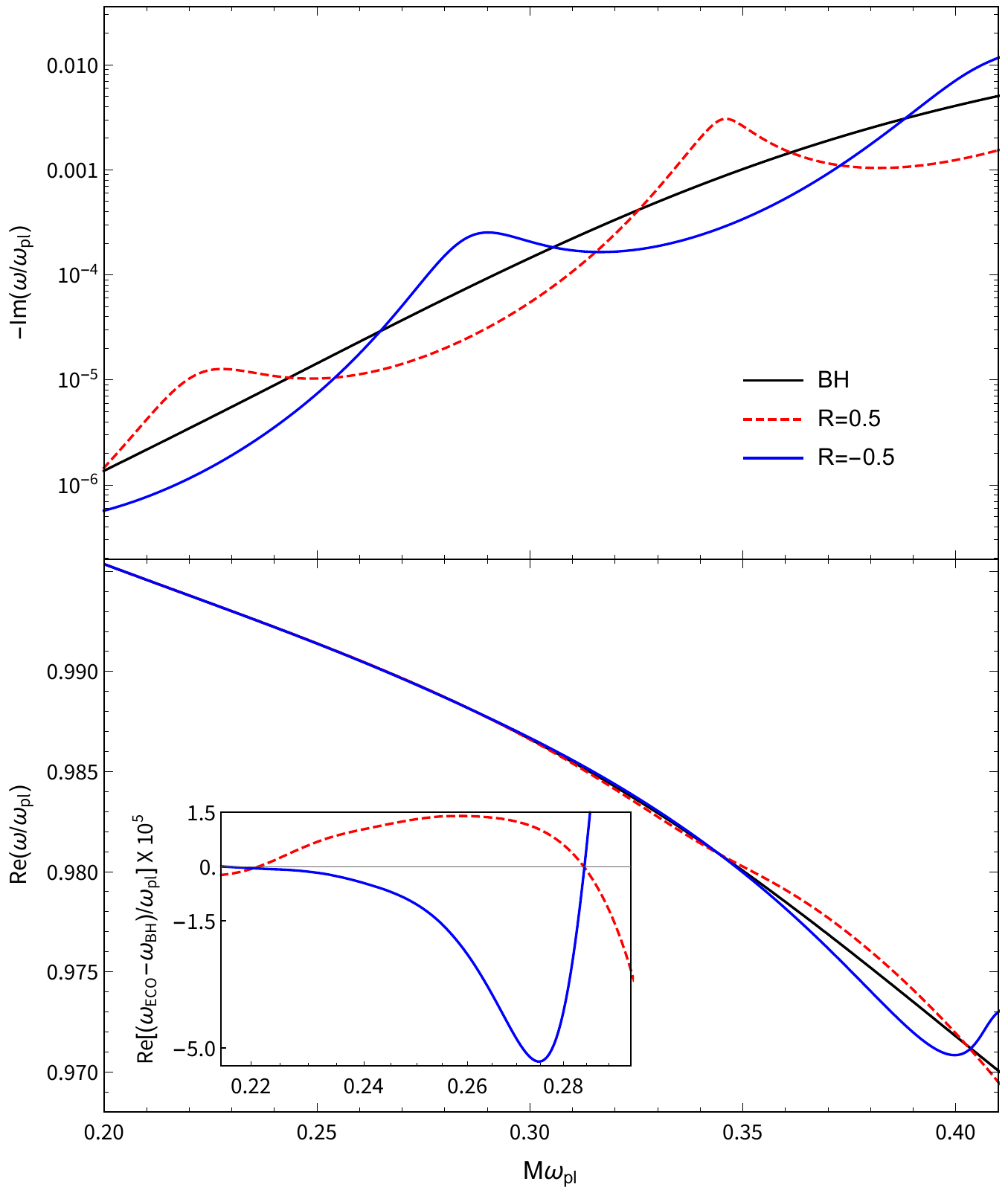}
\caption{Plots of the imaginary and real parts of the frequencies of fundamental quasi-bound states associated with EM waves propagating in the background of Schwarzschild-like ECO in the axial sector, with the plasma frequency $M\omega_{\text{pl}}$ have been presented. For these plots we have chosen $\epsilon=10^{-5}$ and three possible choices of constant reflectivity, $R=0.2, 0.5$ and $-0.5$, respectively. The corresponding frequencies associated with a Schwarzschild BH have also been depicted (the black line in each plot). The left panel shows the frequencies of quasi-bound states for positive reflectivity, whereas the right panel shows quasi-bound state frequencies for both positive and negative reflectivities. The inset in each panel shows the deviation of the real part of the quasi-bound state frequencies from that of a Schwarzschild black hole.}
\label{fig:ReImW-R}
\end{figure*}

Given all these inputs and the numerical method depicted above, the fundamental quasi-bound frequencies for the Schwarzschild black hole (which corresponds to the case $R=0$) and for Schwarzschild-like ECOs have been plotted in \ref{fig:ReImW-R}. We observe that for a fixed value of the reflectivity, both the decay rate (inverse of the imaginary part of the quasi-bound frequency) and the frequency of the quasi-bound state oscillate about the corresponding black hole values. This oscillation is reminiscent of the ``two-well'' structure of the effective potential, one between the maxima of the potential and the surface of the ECO and the other originating from the minima of the effective potential $V_{\rm ax}^{lm}$, which arises due to the effective mass of the photon as it propagates through the plasma. Akin to the QNMs, the relative deviation of the real part of the quasi-bound frequencies for ECOs, from the black hole values, is smaller compared to the corresponding deviation of the decay rate (see \ref{fig:ReImW-R}). Thus, depending on the plasma frequency, $M\omega_{\text{pl}}$, which in turn depends on the electron number density, the quasi-bound states for EM waves around an ECO can be either long-lived or short-lived, compared to a BH. In particular, the deviation of the imaginary part of the quasi-bound states for ECOs from BHs is at most by a factor of 10, as evident from \ref{fig:ReImW-R}, while for QNMs such a deviation is $\mathcal{O}(10^{6})$ and higher depending on the model of ECO. Therefore, in striking contrast to QNMs, the frequencies of quasi-bound states of ECOs are more closer to the corresponding values for BHs. These results are also supported by analytical computations, which have been presented in detail in \ref{app-c}.

Dependence of the quasi-bound frequencies on the reflectivity can also be assessed from \ref{fig:ReImW-R}. As evident, with increasing reflectivity, the deviation of the decay rate from the black hole value increases. For higher values of $M \omega_{\text{pl}}$, the real part of the frequency also shows significant deviation from the black hole results, while the imaginary part shows stronger decay (compared to black holes). Negative values of reflectivity, presented in the second panel of \ref{fig:ReImW-R} imply a phase reversal of the reflected wave from the ECO surface. So far we have limited our analysis for $M\omega_{\rm pl}>0.2$. This is because, extending the results for $M\omega_{\text{pl}}<0.2$ is extremely challenging, particularly for larger reflectivity due to numerical inaccuracies and it becomes exceedingly difficult to track the fundamental mode, mostly because of the oscillatory nature. We would like to emphasize that the above numerical instability arises solely in the case of ECOs, as argued in the previous section.

\begin{figure*}[htbp!]
\includegraphics[width=0.45\textwidth]{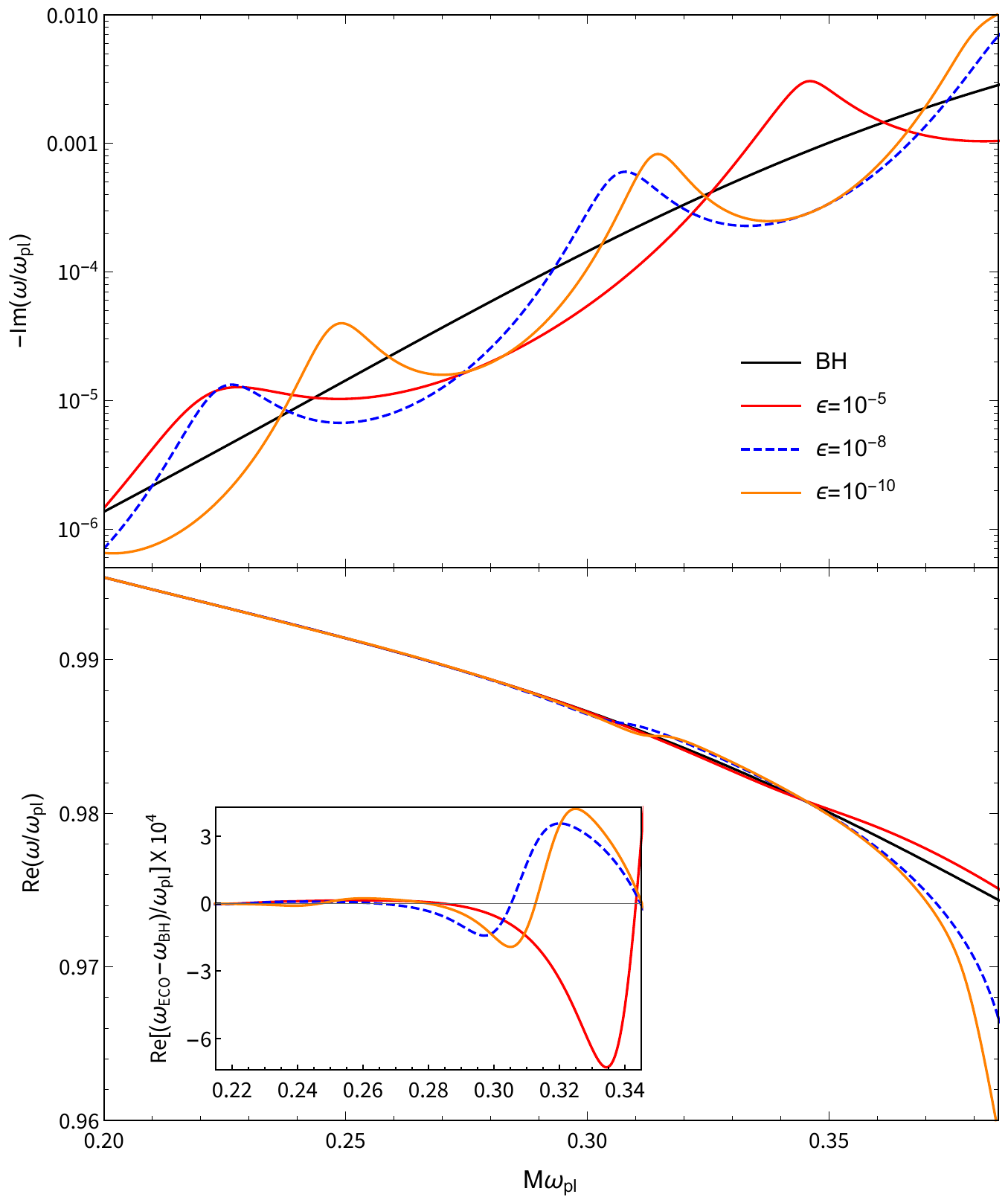}
\includegraphics[width=0.45\textwidth]{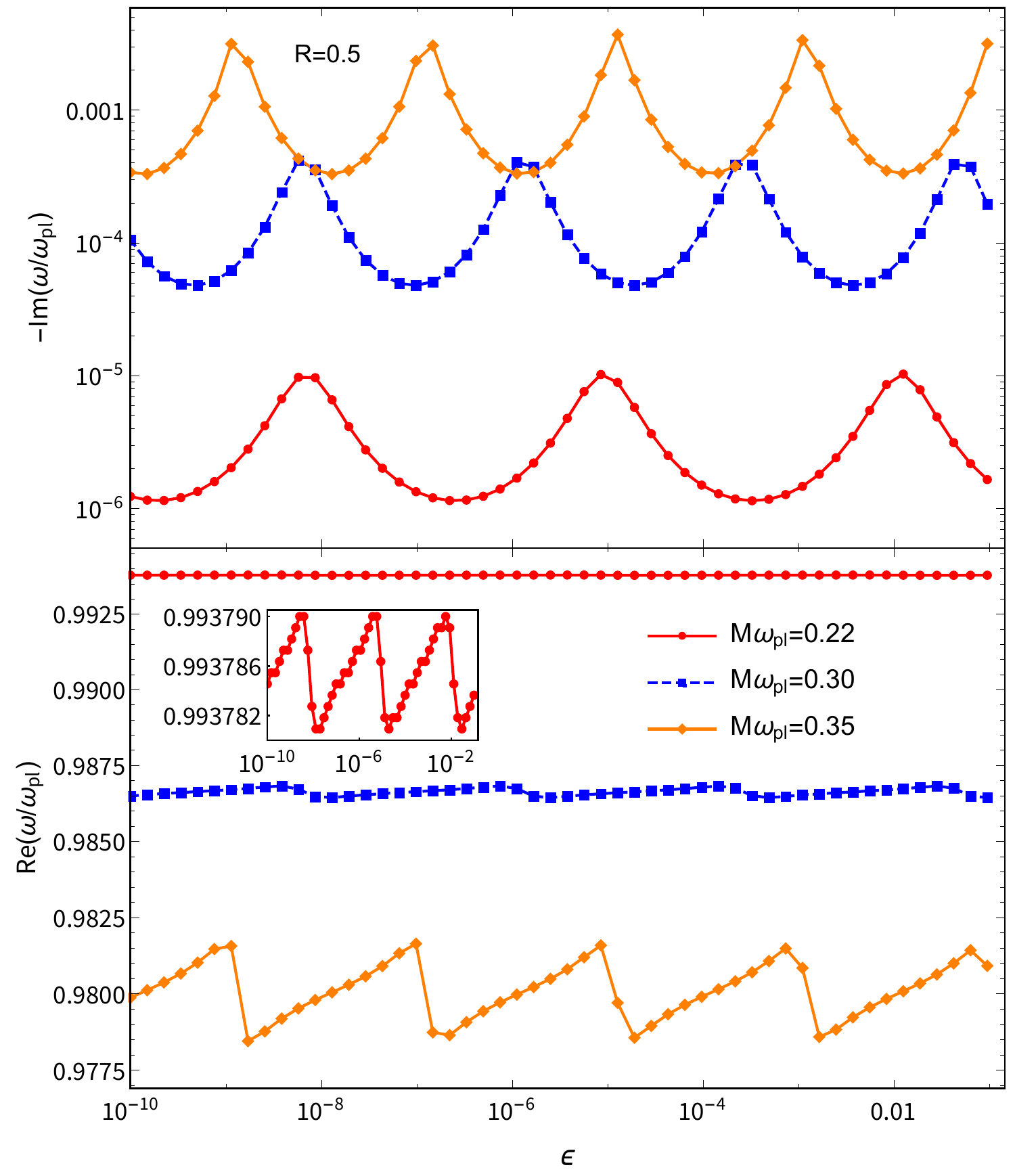}
\caption{Left: plots of the imaginary and real parts of the plasma-driven quasi-bound states frequencies for EM fields associated with a Schwarzschild-like ECO in the axial sector have been presented against the plasma frequency $M\omega_{\text{pl}}$. Here we have fixed the reflectivity to be $R=0.5$, but consider three different positions of the reflective boundary. The inset shows the deviation of the real part of the quasi-bound state frequency from that of a Schwarzschild black hole. Right: plots of the imaginary and real parts of the quasi-bound state frequencies with $\epsilon$ have been presented for fixed reflectivity, $R=0.5$ and three possible choices of the plasma frequency $M\omega_{\rm pl}$. The inset in the lower panel shows the zoomed-in view of the real part of quasi-bound frequencies, for $M\omega_{\text{pl}}=0.22$.}.
\label{fig:ReImW-e}
\end{figure*}

The quasi-bound state frequencies for a given reflectivity also strongly depend on the position of the ECO surface, as can be observed from \ref{fig:ReImW-e}. As the ECO surface moves closer to the maxima of the potential, i.e., as $\epsilon$ increases, the trapping region of the potential in the interior of the maxima becomes smaller (see \ref{fig:pot}), and the oscillation length of the imaginary part of the frequency extends further, covering more regions of the $M\omega_{\text{pl}}$ parameter space. From the right hand side plot of \ref{fig:ReImW-e} it is evident that for larger values of $M\omega_{\rm pl}$, the variation of the real and imaginary parts of the frequencies of quasi-bound states with $\epsilon$ is the strongest, while for smaller values of $M\omega_{\rm pl}$, the variations with $\epsilon$ are weaker. Also the oscillation length of the imaginary part of quasi-bound state frequencies depend on $M\omega_{\rm pl}$, increasing with decreasing values of $M\omega_{\rm pl}$. All of these results are consistent with analytical estimation of the bound state frequencies, presented in \ref{app-c}, showing oscillatory behaviour, as observed in the numerical analysis. 

\begin{figure}[htbp!]
\includegraphics[width=0.99\linewidth]{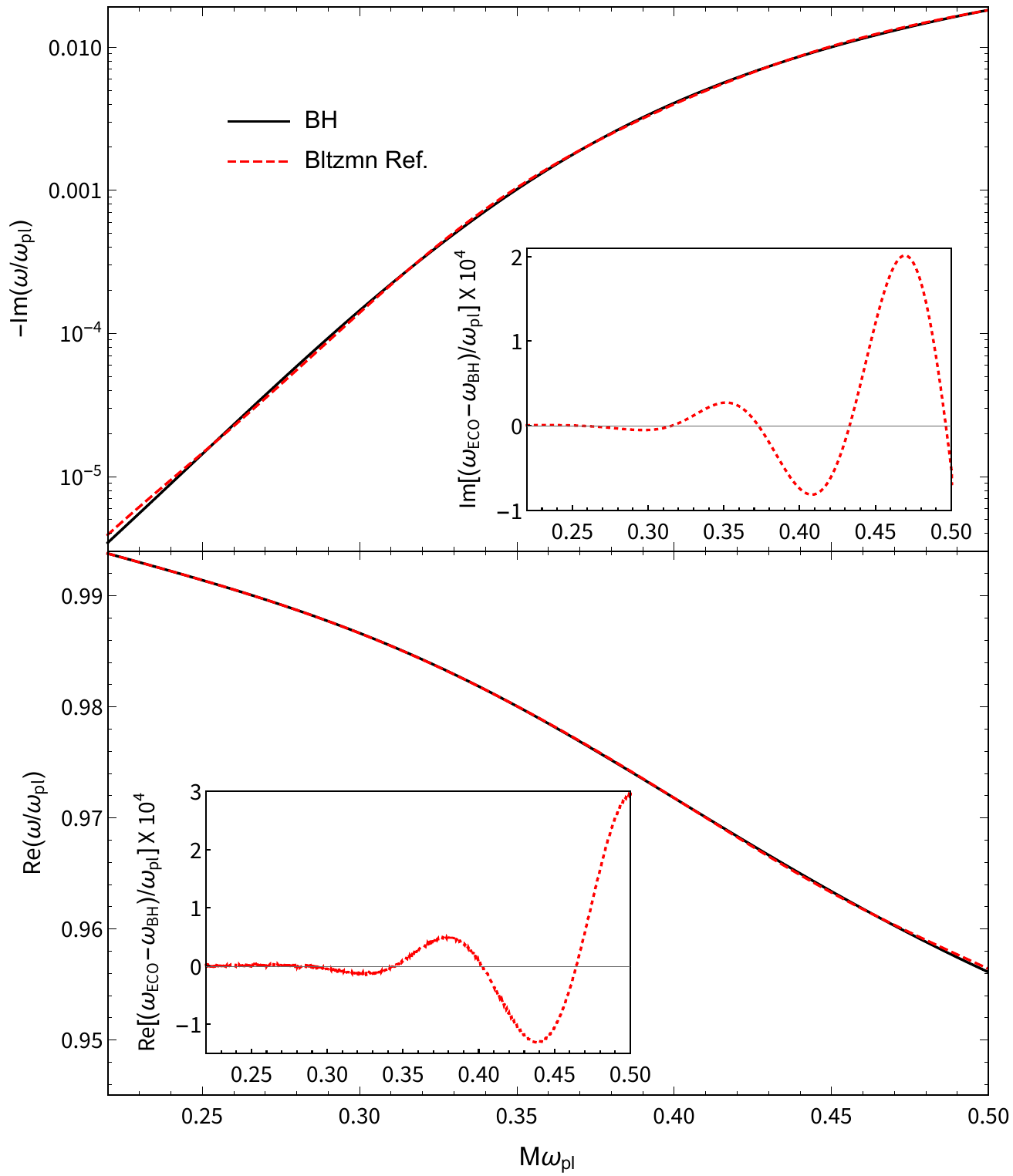}
\caption{Plots of the imaginary and the real parts of the fundamental quasi-bound frequencies for axial EM field around an accreting Schwarzschild-like ECO have been presented, against the plasma frequency $\omega_{\text{pl}}$. We have taken $\epsilon=10^{-5}$, and the reflectivity of the ECO has been taken to be Boltzmann reflectivity. The insets in each plots show the deviation from the black hole results.}
\label{fig:boltzman}
\end{figure}

So far our analysis assumes constant reflectivity, however as emphasized before, the reflectivity of an ECO can in general be frequency dependent. To illustrate implications for such a frequency dependent reflectivity, we consider the case of Boltzmann reflectivity. The Boltzmann reflectivity is akin to any quantum corrected BH, which emits Hawking radiation at temperature $T_{\rm H}$, and since one can assume the BHs to be not perfectly black, rather reflecting as a black body at temperature $T_{\rm H}$, it is natural to assume $R=\exp(-\hbar \omega/k_{\rm B}T_{\rm H})$. The quasi-bound state frequencies for an ECO with Boltzmann reflectivity have been presented in~\ref{fig:boltzman}. As expected, in this context as well, we observe that both the frequency and the decay rate of the plasma-driven EM quasi-bound states oscillate about the black hole values. The deviation from the black hole values increases with increasing $M\omega_{\text{pl}}$ coupling, and hence with increasing the electron number density in the plasma. This is because, increasing $\omega_{\rm pl}$ increases the effective mass of the photon, leading to deeper potential well. This leads to significant deviation from the BH quasi-bound state frequencies.  

\begin{figure}[htbp!]
\includegraphics[width=0.99\linewidth]{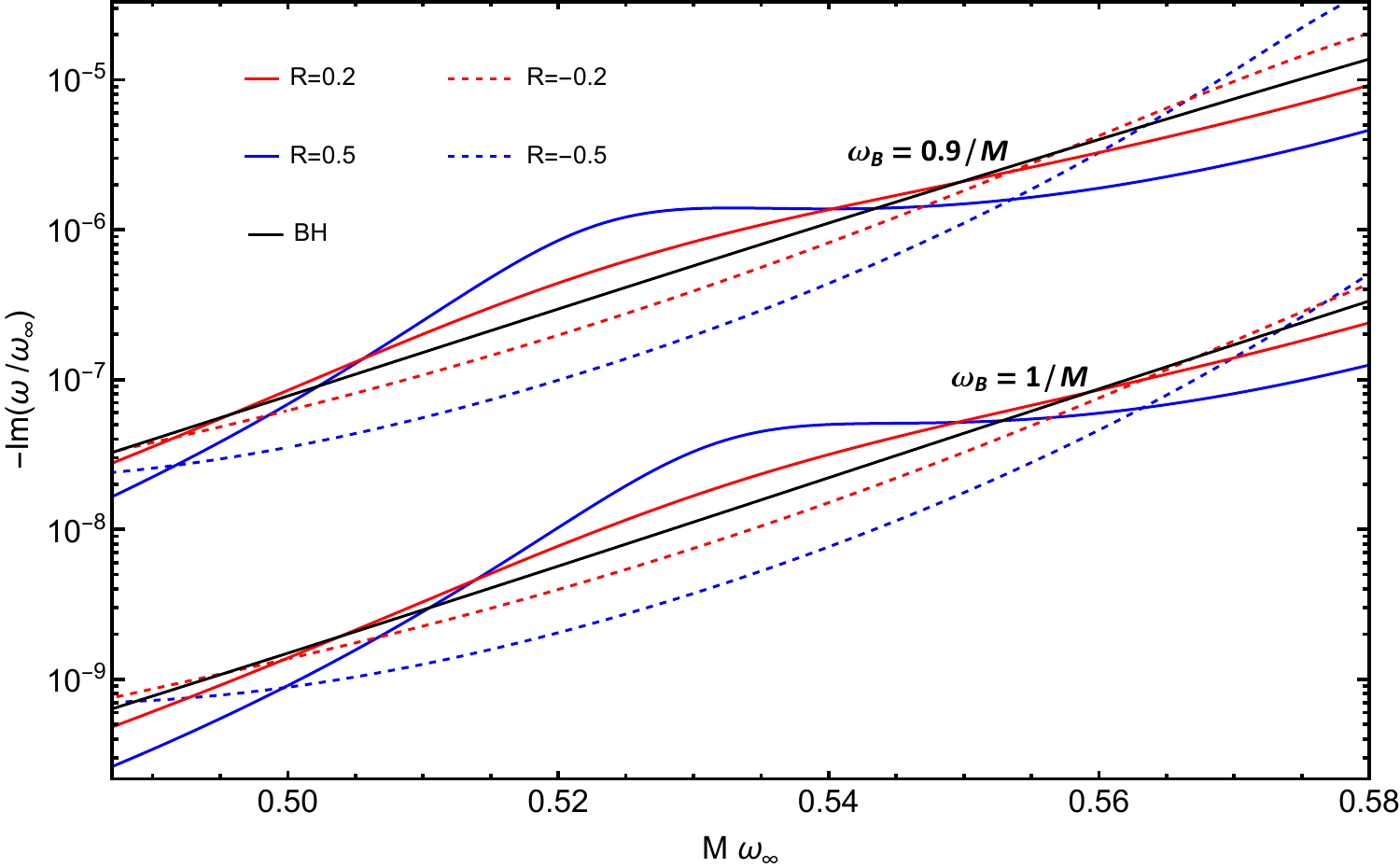}
\caption{Plots of the imaginary part of the plasma-driven quasi-bound frequencies of an axial EM field, in the background of a Schwarzschild-like ECO, has been presented against the asymptotic frequency $\omega_{\infty}$ for Bondi accretion. We have assumed $\epsilon=10^{-5}$ with $\lambda=(3/2)$, i.e., monoatomic gas for the accreting matter. As evident, increasing values for the near-horizon frequency, here we have compared two cases for $\omega_{B}=\frac{1}{M}$ and $\omega_{B}=\frac{0.9}{M}$.}
\label{fig:bondi-odd}
\end{figure}

\subsubsection{Bondi Accretion and quasi-bound state frequencies}

We have described a homogeneous plasma profile and the quasi-bound state frequencies associated with axial EM fields. In this section, we would like to discuss the more realistic Bondi-like accretion, which fits well the spherical symmetry of the background spacetime. The plasma frequency, in the present context is given by \ref{eq:bondi}. In \ref{fig:bondi-odd}, we show the behaviour of the imaginary part of the axial quasi-bound state frequency with $M\omega_{\infty}$ for Bondi accretion. We fix $\lambda=3/2$, corresponding to monoatomic gas. From \ref{fig:bondi-odd} it is clear that as in the homogeneous case, the imaginary part of the quasi-bound state frequency for a given reflectivity oscillates about the monotonically increasing BH results. The real part of the quasi-bound state frequency behaves similarly as in the homogeneous distribution, however, the deviation from the BH result is extremely small and hence has not been depicted here. As in the BH case~\cite{cannizzaro2021PRD}, for ECOs as well, the decay rate of the quasi-bound state around a Schwarzschild-like ECO decreases with increasing the surface plasma frequency $\omega_{\rm B}$. The corresponding deviation from the BH results, on the other hand, decreases with increasing $\omega_{\rm B}$. Thus, unlike BHs, for certain choices of $\omega_{\rm B}$, these quasi-bound states associated with ECOs decay sufficiently slowly, making these plasma frequencies susceptible to instabilities. Again this is a feature unique to the ECOs. In general, for a given surface plasma frequency $\omega_{\rm B}$, the imaginary part of the quasi-bound state frequency is smaller compared to the homogeneous plasma distribution. This makes it more susceptible to numerical instability.

\subsection{Quasi-bound states: Polar sector}\label{sec:3c}

\begin{figure}[htbp!]
\includegraphics[width=0.99\linewidth]{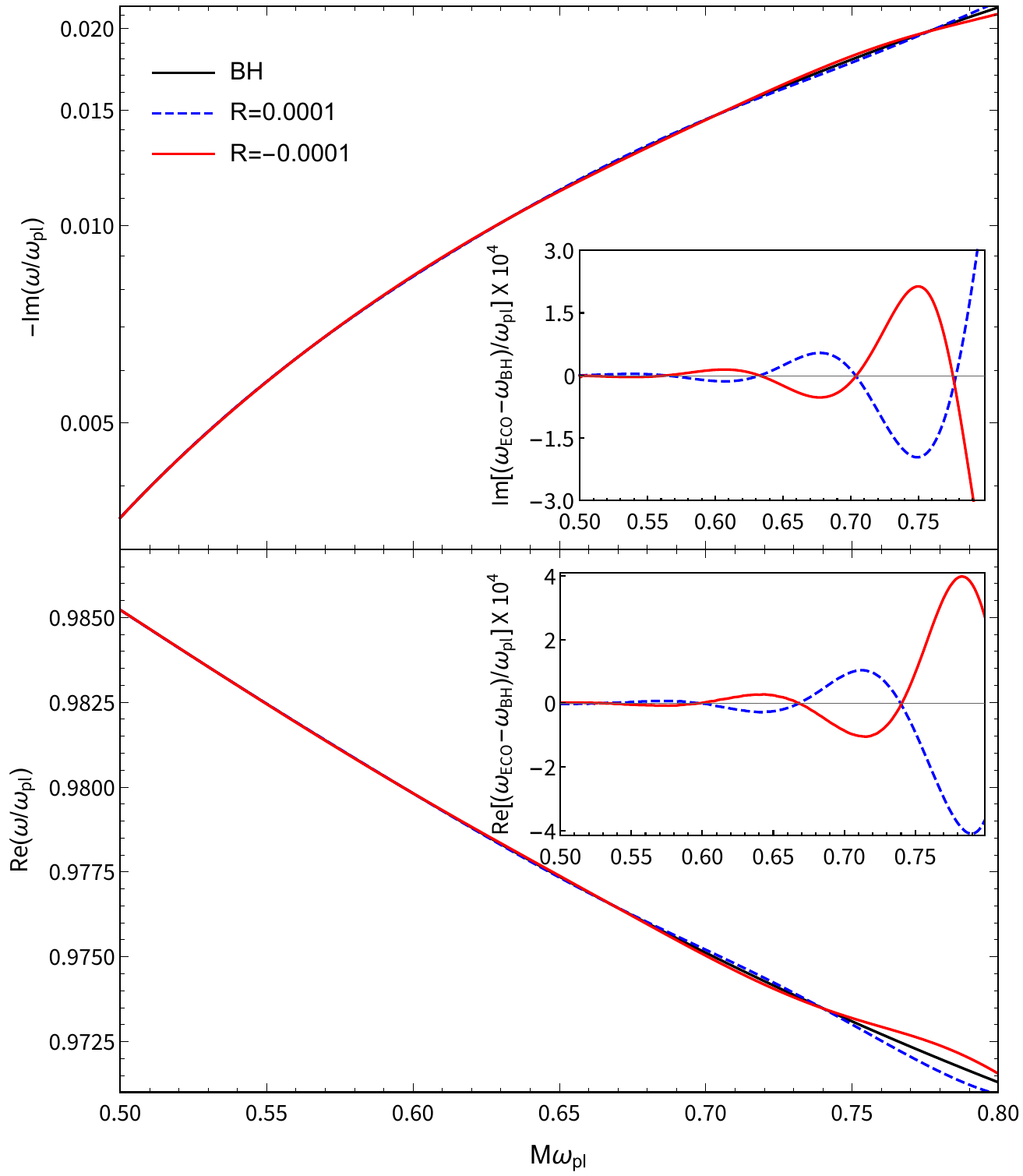}
\caption{The imaginary and the real parts of the fundamental plasma-driven quasi-bound frequencies have been depicted, against the plasma frequency, for the polar sector of an EM wave moving in the background of a Schwarzschild-like ECO. We have taken the compactness of the ECO, such that $\epsilon=10^{-5}$, and the reflectivity is taken to be $R=\pm 0.0001$. The black curve corresponds to the frequencies associated with Schwarzschild BH. The inset in each plot shows the corresponding deviation from the BH results.}
\label{fig:ReIm-even}
\end{figure}

In this final section we discuss the frequencies of quasi-bound states associated with the polar sector. The main difficulty with the polar sector is the complicated structure of the effective potential, presented in \ref{eq:pl-even}, and in particular, the nontrivial dependence of the effective potential on $\omega$. Therefore, we work exclusively on the homogeneous plasma distribution and depict the variation of the plasma-driven EM quasi-bound frequencies with $M\omega_{\text{pl}}$ in \ref{fig:ReIm-even}.  

We note that, as in the axial case, both the frequency and the decay rate oscillate about the BH result, and the amplitude of oscillation increases with $M\omega_{\text{pl}}$. This is due to deepening of the potential well with increasing plasma frequency. Due to the nontrivial dependence of the effective potential in the polar sector on $\omega$, the numerical evolution becomes increasingly unstable as the reflectivity is increased, and hence we had to work with small reflectivities. The general behaviour of the quasi-bound state frequencies and the decay rates remain similar to that in the axial sector. For Boltzmann-reflectivity, on the other hand, the quasi-bound state frequencies shows similar behaviour as in the axial sector, however, since the amplitude of deviation from the BH results is even smaller it has not been depicted here. At this outset, it must be emphasized that the behaviour of the quasi-bound state frequencies associated with the polar sector are similar to that of the axial sector. The smallness in the difference of the quasi-bound frequencies from the BH values is due to the small reflectivity for which stable numerical results can be derived in the polar sector.

For the Bondi accretion, with inhomogeneous plasma frequency, the effective potential in the polar sector becomes even more complicated and the numerical accuracy becomes worse. The small values of the imaginary part of the quasi-bound state frequencies add up to the difficulty. 

As evident from \ref{fig:ReIm-even}, even in the polar sector, depending on the value of the plasma frequency, the quasi-bound states associated with ECO either decay with a longer or, shorter time scale compared to those around a BH. In particular, the oscillating pattern around the BH value is seen in the polar sector as well. It is important to comment that throughout our analysis we have not observed any unstable modes, i.e., modes with positive imaginary parts. Hence within the purview of our analysis, the plasma-driven EM quasi-bound states around static and spherically symmetric ECOs are stable, but the degree of stability is weaker or stronger compared to the corresponding BH.

\section{Conclusion}\label{sec:4}

Proving the existence of BHs and exploring its properties has been one of the pinnacle for research on gravitational physics. With the detection of GWs by the LVK collaboration, we are merely one step away from unveiling the true nature of BHs, which are abundant in our universe, from the center of our galaxies to various astrophysical scenarios. In order to really test the BH hypothesis, we need alternatives to BHs, and these objects are collectively dubbed as ECOs, which are horizonless but possess a light ring similar to that of a BH. These ECOs are characterized by a real surface with a radius greater than the event horizon of a corresponding BH. As far as an external observer is concerned, the interior properties of an ECO are represented by two physical quantities --- (a) the reflectivity $R$ of the ECO surface, and (b) the location of its surface, characterized by $\epsilon$. There have been plethora of works, discussing various properties of these ECOs, but mostly in the absence of any environment. In the present work, we have investigated the effect of accretion onto the stability of such an ECO. In detail, we have studied the propagation of EM waves through a cold and collisionless plasma accreting onto a Schwarzschild-like ECO. The symmetry of the background geometry allowed us to decompose the evolution equations for the EM field in axial and polar sectors, which decouples from one another. We have presented, for the first time, the equations governing the dynamics of the EM wave propagating in a plasma within a general static and spherically symmetric background, both for the polar and the axial sectors. Though, in the axial sector the governing dynamical equations are identical to that of a Proca field, in the polar sector, the dynamical equations for the EM field are completely different. This observation matches with~\cite{cannizzaro2021PRD}. 

Subsequently, we have applied our general equations to study the quasi-bound states of the EM waves and their stability for Schwarzschild-like ECO. In particular, our interest lies in understanding the effect of reflectivity and the location of the ECO surface on the frequencies of the plasma-driven quasi-bound states. These states are particularly interesting in a rotating compact object as they are prone to superradiant instability. Of course, a spherically symmetric compact object is not subjected to superradiance but a study of the quasi-bound states even in such a system may provide valuable insight into the stability of such compact objects.

To study the frequency spectrum of quasi-bound states, we employ the direct integration method with reflective boundary conditions near the surface of the ECO and exponentially decaying behaviour asymptotically. In all the cases studied, we observed the general trend that the decay rate of the quasi-bound states increases with the plasma frequency $M \omega_{\text{pl}}$, whereas the real frequency of the quasi-bound state decreases with the same. For a given reflectivity both the real and the imaginary parts of the quasi-bound state frequencies oscillates about the BH results. The amplitude of oscillation increases with the reflectivity. On the other hand, as the compactness of the ECO increases (smaller values of $\epsilon$), the oscillation length of the frequencies for quasi-bound states decreases. Therefore, the $M\omega_{\text{pl}}$ parameter space gets populated with more short and long lived modes compared to BHs. Thus, for a Schwarzschild-like ECO with a given mass, depending on the electron number density, the quasi-bound states may be short or long-lived compared to a BH. This is true for both homogeneous and Bondi-like plasma distribution, with constant as well as frequency-dependent reflectivity, in both the polar and axial sectors. It is important to note that the relative deviation of the real part of the quasi-bound frequencies of the ECO from the BH quantities are much smaller than the corresponding deviation in the decay rate. Moreover, the axial modes are short-lived compared to the polar modes, in general. All of these numerical results are consistent with our analytical findings presented in \ref{app-c}. Finally, we have also observed that, if the reflectivity of the ECO surface is frequency-dependent, as in the case of Boltzmann reflectivity, the deviation from the BH result decreases. 

Through our analysis we were able to show that, unlike the QNM frequencies, the frequencies of quasi-bound states of EM waves associated with ECOs differs from the BH values by an order or two. However, the frequencies of quasi-bound states around ECOs describe a unique oscillation pattern around BH values, whose amplitude is related to the reflectivity of the ECO, and the oscillation length is dependent on $\epsilon$ --- a new feature for ECOs. Our analyses enabled us to explore a considerable portion of the parameter space within which, we did not find any unstable modes indicating the stability of the ECOs under plasma driven EM perturbation. The results also strongly indicate that in the case of rotating ECOs, even in the linear regime, the presence of plasma will result in stronger quenching (compared to black holes) of superradiant instability for a certain part of the parameter space, where the imaginary part of quasi-bound state frequency is larger compared to a BH, while for some other parameters the superradiant instability will be stronger for ECOs. Thus, it will be interesting to study how the presence of plasma affects the instability time scale in the case of rotating ECOs~\cite{PhysRevD.108.103025}, which we hope to address in a separate work. Along these lines, it will also be interesting to study the effect of nonlinearity and magnetic field in the photon-plasma interaction on the background of ECOs. Similar studies are already underway for black holes~\cite{Cannizzaro:2021zbp, PhysRevD.109.023007, Spieksma:2023vwl} and we hope to report the consequences for ECOs elsewhere. 

\begin{acknowledgements}
SB thanks IACS for financial support and Somsubhra Ghosh and Pratick Sarkar for some helpful discussion. AC and SC acknowledge hospitality from IIT Gandhinagar and IUCAA, where a substantial part of the work was done. The authors acknowledge Surajit Ghosh for useful discussions. AC acknowledges financial support from SERB, Government of India via the National Postdoctoral Fellowship (File No. PDF/2023/000550).  Research of SC is supported by the Mathematical Research Impact Centric Support (MATRICS) and the Core research grants from SERB, Government of India (Reg. Nos. MTR/2023/000049 and CRG/2023/000934).
\end{acknowledgements}
\appendix
\labelformat{section}{Appendix~#1}
\onecolumngrid
\section{Effective potential in the Polar sector}\label{app:a}

In this appendix we will present the full expression for the effective potential in the polar sector. The expression involves the metric function, as well as the plasma frequency and their derivatives, which is given by 
\begin{align}\label{eq:V-pl-even}
V^{lm}_{\text{pol}}&=\frac{1}{4 r^2 \Bigl(\omega^2 -  f \omega_{\text{pl}}^2\Bigr)^2 \Bigl[r^2 \omega^2 -  f \Bigl\{l(l+1)+r^2\omega_{\text{pl}}^2\Bigr\}\Big]^2}
\Bigg[-r^4 \omega^6 \Bigl\{4 r^2 \omega^4 -  l (l+1) g f'^2\Bigr\}
+ r^2 \omega^4 f \Biggl\{2 r^2 \omega_{\text{pl}}^2 \Bigl(10 r^2 \omega^4
\nonumber
\\
&\qquad+l(l+1)gf'^2\Bigr)+l(l+1)\biggl[r^2 \omega^2 \bigl(12 \omega^2 + f' g'\bigr)
+ 2 g \Bigl(-5 r \omega^2 f' + l (l + 1) f'^2 + r^2 \omega^2 f''\Bigr)\biggr]\Biggr\}
\nonumber
\\
&\qquad - r\omega^2 f^2 \Bigg\{r \omega^2 \biggl(40 r^4 \omega^2 \omega_{\text{pl}}^4+ 2 l (l +1) r^2 \omega_{\text{pl}}^2 \bigl(24 \omega^2 + f' g'\bigr) + l (l + 1) \Bigl[12 l (l + 1) \omega^2 + \bigl(2 r \omega^2 + l (l + 1) f'\bigr) g'\Bigr]\biggr)
\nonumber
\\
&\qquad + l (l + 1) g \Biggl(3 r \omega_{\text{pl}}^2 \Bigl[l (l + 1) + r^2 \omega_{\text{pl}}^2\Bigr] f'^2 
+ 2 \omega^2 f' \Bigl[l (l + 1) - 11 r^2 \omega_{\text{pl}}^2 - 9 r^3 \omega_{\text{pl}} \omega'_{\text{pl}}\Bigr]
+ 2 r \omega^2 \biggl[-6 \omega^2
\nonumber
\\
&\qquad + \Bigl\{l (1 + l) + 2 r^2 \omega_{\text{pl}}^2\Bigr\} f''\biggr]\Biggr)\Bigg\} 
+ 2 f^5 \omega_{\text{pl}}^3 \Biggl\{6 l (l + 1) r^4 \omega_{\text{pl}}^5 + 2 r^6 \omega_{\text{pl}}^7 
+ l (l + 1) r^2 \omega_{\text{pl}}^3 \Bigl(6 l (l + 1) - 6 g + r g'\Bigr) 
\nonumber
\\
&\qquad + l (l + 1) \omega_{\text{pl}} \biggl(l (l + 1) r g' + 2 \Bigl\{l^2 (l + 1)^2 - 3 r^4 g f'^2\Bigr\}\biggr)
+l^2(l + 1)^2 r \Bigl\{r g' \omega'_{\text{pl}} + 2 g \bigl(2 \omega'_{\text{pl}} + r \omega''_{\text{pl}}\bigr)\Bigr\}
\nonumber
\\
&\qquad + l(l + 1) r^3 \omega_{\text{pl}}^2 \Bigl[r g' \omega'_{\text{pl}} + g \bigl(-8 \omega'_{\text{pl}} + 2 r \omega''_{\text{pl}}\bigr)\Bigr]\Biggr\}
\nonumber
\\ 
&\qquad - 2 f^4 \Biggl\{10 r^6 \omega^2 \omega_{\text{pl}}^8 + l (l + 1) r^3 \omega_{\text{pl}}^6 \bigl(24 r \omega^2 -  g f'\bigr) + l (l + 1) r \omega_{\text{pl}}^4 \Bigl[- g \bigl(18 r \omega^2 + l (l+1) f'\bigr)+ 3 r \omega^2 \bigl(6 l (l + 1) + r g'\bigr)\Bigr]
\nonumber
\\
&\qquad -l (l + 1) r^4 g \omega_{\text{pl}}^5 f' \omega'_{\text{pl}} + 2 l^2 (l + 1)^2 r^2 \omega^2 g f'^2 + 2 l (l + 1) \omega^2 \omega_{\text{pl}}^2 \biggl(l (l + 1) r g' + 2 \Bigl\{l^2 (l + 1)^2 -  r^4 g f'^2\Bigr\}\biggr) 
\nonumber
\\
&\qquad + l^2 (l + 1)^2 r \omega^2 \omega_{\text{pl}} \Bigl\{r g' \omega'_{\text{pl}} + 2 g \bigl(2 \omega'_{\text{pl}} + r \omega''_{\text{pl}}\bigr)\Bigr\}
-l (l + 1) r^2 \omega_{\text{pl}}^3 \biggl(-2 r^2 \omega^2 g' \omega'_{\text{pl}} 
+ g \Bigl[\bigl\{16 r \omega^2+ l (l + 1) f'\bigr\}\omega'_{\text{pl}}
\nonumber
\\
&\qquad -4 r^2 \omega^2 \omega''_{\text{pl}}\Bigr]\biggr)\Biggr\}
+ \omega^2 f^3 \Biggl\{40 r^6 \omega^2 \omega_{\text{pl}}^6 - 20 l (l + 1) r^4 g \omega_{\text{pl}}^3 f' \omega'_{\text{pl}}
+ 2 l (l + 1) \omega^2 \biggl(l (l + 1) r g' + 2 \Bigl[l^2 (l + l)^2 + r^4 g f'^2\Bigr]\biggr)
\nonumber
\\ 
&\qquad+l (l + 1) r^3 \omega_{\text{pl}}^4 \bigl(72 r \omega^2 - 14 g f' + r f' g'+ 2 r g f''\bigr) + l (l + 1) r^2 \omega_{\text{pl}}^2 \Bigl[36 l (l + 1) \omega^2 + \bigl(6 r \omega^2 + l (l + 1) f'\bigr) g' 
\nonumber
\\
&\qquad+g \bigl(-36 \omega^2 + 2 l (l + 1) f''\bigr)\Bigr] - 2 l (l + 1) r^2 \omega_{\text{pl}} \biggl(- r^2 \omega^2 g' \omega'_{\text{pl}} + g \Bigl[\bigl\{8 r \omega^2 + 3 l (l + 1) f'\bigr\} \omega'_{\text{pl}} - 2 r^2 \omega^2 \omega''_{\text{pl}}\Bigr]\biggr)\Biggr\}\Bigg]~.
\end{align}
This is the expression we have used in our numerical analysis to determine the frequencies of the quasi-bound states.

\section{Relative location of maxima and minima for the axial effective potential}\label{app-b}

The axial effective potential $V_{\rm ax}^{lm}$ is given by \ref{eq:V-pl-odd}, which admits a minima and a maxima. The maxima in the potential is akin to the light ring, while the minima specifically originates from the effective mass associated with the axial sector of the EM field. For existence of quasi-bound orbits it is necessary that the minima lies at a larger radial coordinate than the maxima, which gies a condition on the parameters of the problem. Setting ${V_{\rm ax}^{lm}}'=0$, we obtain the following algebraic equation, 
\begin{align}\label{extrema-condition}
r^{2}-\frac{l(l+1)}{M\omega^{2}_{\text{pl}}}r+\frac{3l(l+1)}{\omega^{2}_{\text{pl}}}=0~,
\end{align}
whose solutions depict the locations of the extrema of the potential ${V_{\rm ax}^{lm}}$. As evident from the above equation, both the roots will exist iff the following condition holds,
\begin{align}
M\omega_{\text{pl}}\leq \sqrt{\frac{l(l+1)}{12}}~.
\end{align}
The above inequality suggests that for $l=1$, for the existence of double root, we must have $M\omega_{\text{pl}}\leq (1/\sqrt{6})$. When this condition is satisfied, the two roots of the above algebraic equation read,
\begin{align}
r_{>}=&\frac{1}{2}\left[\frac{l(l+1)}{M\omega_{\text{pl}}^{2}}+\sqrt{\frac{l(l+1)}{\omega_{\text{pl}}^{2}}\left[\frac{l(l+1)}{M^{2}\omega_{\text{pl}}^{2}}-12\right]}\right]~,
\\
r_{<}=&\frac{1}{2}\left[\frac{l(l+1)}{M\omega_{\text{pl}}^{2}}-\sqrt{\frac{l(l+1)}{\omega_{\text{pl}}^{2}}\left[\frac{l(l+1)}{M^{2}\omega_{\text{pl}}^{2}}-12\right]}\right]~.
\end{align}
One can easily check that $r_{>}>r_{<}$, along with the fact that $r_{>}$ is the location of the minima, while $r_{<}$ is the location of the maxima. Implying that the minima is always at a larger distance than the maxima. It can also be verified by computing the potential at the location of extrema, which reads,
\begin{align}\label{pot-at-extrima}
V(r)=\omega_{\text{pl}}^{2}+\left[\frac{4Ml(l+1)}{r^{3}}-\frac{l(l+1)}{r^{2}}\right]~.
\end{align}
such that,
\begin{align}
V(r_{<})- V(r_{>})=\frac{(r_{>}-r_{<})l(l+1)}{r_{>}^{2}r_{<}^{2}}\left[\frac{l(l+1)}{3M\omega_{\text{pl}}^{2}}-4M\right]\geq 0~.
\end{align}
Therefore, considering the monopole mode, as long as $M\omega_{\text{pl}}\leq (1/\sqrt{6})\simeq 0.4$, there exists a trapping well outside the light ring, resulting into the formation of quai-bound states. This result has been used in the main text. 

\section{Analytical estimate of the axial quasi-bound frequencies for Schwarzschild-like ECO}
\label{app-c}

In this appendix, we will provide analytical results demystifying the oscillatory behaviour of the axial quasi-bound frequencies associated with ECOs, derived through numerical methods. For this purpose, we start with the following equation governing the propagation of axial EM wave,
\begin{align}\label{Eq-ax-as-per}
x^{2}(1+x)^{2}{\Psi_{\text{ax}}^{lm}}''+x(1+x){\Psi_{\text{ax}}^{lm}}'+\left[\omega^{2}r^{2}_{+}(1+x)^{4}-x(1+x)\left\{\omega^{2}_{\text{pl}}r_{+}^{2}(1+x)^{2}+l(l+1)\right\}\right]\Psi_{\text{ax}}^{lm}=0~,
\end{align}
where, we have introduced the dimensionless radial coordinate $x\equiv(r/r_{+})-1$, and `prime' denotes derivative with respect to $x$. In what follows we will divide the spacetime into three regions --- (i) the near horizon region (or, near ECO region) defined as $(M\omega_{\text{pl}})x\ll1$ and $(M\omega)x\ll1$, (ii) the far region, which is defined as $x\gg 1$, and most importantly, (iii) the intermediate region, defined as $1\ll x \ll (l/M\omega)$. We will solve the above equation analytically in the near horizon and far region, and shall match the solutions in the intermediate region to determine the frequencies. 

Near the horizon, using the approximations discussed above, \ref{Eq-ax-as-per} takes the following form,
\begin{align}\label{near-hor-axial-eq}
x^{2}(1+x)^{2}{\Psi^{lm}_{\text{ax}}}''+x(x+1){\Psi^{lm}_{\text{ax}}}'+\left[\omega^{2}r_{+}^{2}-x(x+1) l(l+1)\right]\Psi_{\text{ax}}^{lm}=0~.
\end{align}
For simplicity, we will henceforth set $M=1$, and the solution to the above equation takes the following form, 
\begin{align}\label{Near-wall-eq-sol}
\Psi^{lm~\text{(near)}}_{\text{ax}}&=A\, x^{-2i\omega}(1+x)^{1+\delta}\,{}_{2}F_{1}(-l-2i\omega+\delta,l+1-2i\omega+\delta,1-4i\omega,-x)
\nonumber
\\
&\qquad +B\, x^{2i\omega}(1+x)^{1+\delta}\,{}_{2}F_{1}(-l+2i\omega+\delta,l+1+2i\omega+\delta,1+4i\omega,-x)~,   
\end{align}
where $\delta\equiv \sqrt{1-4\omega^{2}}$, with $A$ and $B$ being arbitrary constants to be fixed later by imposing appropriate boundary conditions. One such boundary condition is imposed by taking $x\to 0$ limit of the above near horizon solution, which takes the following form, 
\begin{align}
\Psi_{\text{ax}}^{lm~\text{(near)}}(x\to 0) \simeq A e^{-i\omega r_{*}}+ B e^{+i\omega r_{*}}~,
\end{align}
and then realizing that the ECO surface is reflective in nature. This allows us to relate the ratio $(B/A)$ to the reflectivity of the ECO surface as in \ref{def_reflect}. 

Further, using the asymptotic behaviour of the hypergeometric function, we can show that in the intermediate region the near horizon solution takes the following form,
\begin{align}\label{Near-intermidiate-sol}
\!\!\!\!\!\! \Psi_{\text{ax}}^{lm~\text{(near)}}\left(x\rightarrow x_{\text{int}}\right)&\simeq \Bigg[A\left(\frac{\Gamma(1-4i\omega)\Gamma(2l+1) }{\Gamma(l+1-2i\omega+\delta)\Gamma(l+1-2i\omega-\delta)}\right) +B\left(\frac{\Gamma(1+4i\omega)\Gamma(2l+1) }{\Gamma(l+1+2i\omega+\delta)\Gamma(l+1+2i\omega-\delta)}\right)\Bigg]x^{l+1}
\nonumber
\\
&+\Bigg[A\left(\frac{\Gamma(1-4i\omega)\Gamma(-2l-1)}{\Gamma(-l-2i\omega+\delta)\Gamma(-l-\delta-2i\omega)}\right)+B\left(\frac{\Gamma(1+4i\omega)\Gamma(-2l-1)}{\Gamma(-l+2i\omega+\delta)\Gamma(-l+2i\omega-\delta)}\right)\Bigg]x^{-l}~.
\end{align}
Here, $1\ll x_{\rm int}\ll (l/\omega)$. As evident, the part of the solution in the first line denotes the growing mode, while the part in the second line depicts the decaying mode. 
Along identical lines, in the far region, the dynamical equation for the axial sector of the plasma-driven EM modes becomes,
\begin{align}\label{Far-with-appx-Scheme}
x^{2}{\Psi_{\text{ax}}^{lm}}''+\left[-q^{2}x^{2}+\nu q x-l(l+1)\right]\Psi_{\text{ax}}^{lm}=0~;
\quad 
q\equiv-\sqrt{\omega_{\text{pl}}^{2}-\omega^{2}}~;
\quad 
\nu\equiv\frac{2\omega^{2}-\omega_{\text{pl}}^{2}}{q}~.
\end{align}
The general solution of the above differential equation is given by a linear combination of confluent hypergeometric functions, which read,
\begin{align}\label{Far-general-sol}
\Psi_{\text{ax}}^{lm~\text{(far)}}=e^{-2qx}\left[C\,(4qx)^{l+1}\, U(l+1-\nu,2l+2,4qx)+D\, (4qx)^{-l}\, M(-l-\nu,-2l,4qx)\right]~.
\end{align}
For depicting a bound state, we demand the above solution to be an exponentially decaying one near infinity, and using the properties of confluent hypergeometric function one can show that the above condition will hold true if and only if, we choose $D=0$. With this choice, in the intermediate regime, we have the following behaviour for the far region solution,
\begin{align}\label{Far-intermidiate-sol}
\Psi_{\text{ax}}^{lm~\text{(far)}}\left(x\rightarrow x_{\text{int}}\right)\simeq C\frac{(4q)^{l}\pi}{\sin[2(l+1)\pi]}\left[\frac{x^{l+1}}{\Gamma(-l-\nu)\Gamma(2l+2)}-\frac{(4q)^{-(2l+1)}x^{-l}}{\Gamma(l+1-\nu)\Gamma(-2l)} \right]~.  
\end{align}
This also has a growing and a decaying mode. Therefore comparing the growing and the decaying modes in \ref{Near-intermidiate-sol} with the corresponding modes in \ref{Far-intermidiate-sol}, and eliminating the arbitrary constant $C$, we arrive at the following relation between $A$ and $B$, 
\begin{align}\label{B-by-A}
(4q)^{2l+1}\frac{\Gamma(-2l-1)\Gamma(-2l)\Gamma(l+1-\nu)}{\Gamma(2l+1)\Gamma(2l+2)\Gamma(-l-\nu)}=-\frac{\frac{\Gamma(1-4i\omega)}{\Gamma(l+1-2i\omega+\delta)\Gamma(l+1-2i\omega-\delta)} +\left(\frac{B}{A}\right)\frac{\Gamma(1+4i\omega)}{\Gamma(l+1+2i\omega+\delta)\Gamma(l+1+2i\omega-\delta)}}{\frac{\Gamma(1-4i\omega)}{\Gamma(-l-2i\omega+\delta)\Gamma(-l-\delta-2i\omega)}+\left(\frac{B}{A}\right)\frac{\Gamma(1+4i\omega)}{\Gamma(-l+2i\omega+\delta)\Gamma(-l+2i\omega-\delta)}}~.
\end{align}
The above condition can be written in a more compact form as,
\begin{align}\label{BS-condition-rearranged}
\frac{\Gamma(-l-\nu)\Gamma(2l+2)}{\Gamma(-2l)\Gamma(l+1-\nu)}=-(4q)^{2l+1}\frac{\Gamma(-2l-1)}{\Gamma(2l+1)}\frac{\Gamma(l+1-2i\omega+\delta)\Gamma(l+1-2i\omega-\delta)}{\Gamma(-l-2i\omega+\delta)\Gamma(-l-\delta-2i\omega)}\frac{1}{\mathcal{G}}~,
\end{align}
where $\mathcal{G}$ has the following form,
\begin{align}\label{def-G}
\mathcal{G}=\frac{1+\frac{B}{A}\frac{\Gamma(1+4i\omega)}{\Gamma(1-4i\omega)}\frac{\Gamma(l+1-2i\omega+\delta)}{\Gamma(l+1+2i\omega+\delta)}\frac{\Gamma(l+1-2i\omega-\delta)}{\Gamma(l+1+2i\omega-\delta)}}{1+\frac{B}{A}\frac{\Gamma(1+4i\omega)}{\Gamma(1-4i\omega)}\frac{\Gamma(-l-2i\omega+\delta)}{\Gamma(-l+2i\omega+\delta)}\frac{\Gamma(-l-2i\omega-\delta)}{\Gamma(-l+2i\omega-\delta)}}~.
\end{align}
It is clear that the contribution of the reflective nature of the ECO surface comes from the $\mathcal{G}$ term. If we had a BH instead, then $B=0$ and hence we have $\mathcal{G}=1$. In this case, the above expression matches with the one of \cite{Rosa:2011my}. In order to simplify it further, we rewrite \ref{BS-condition-rearranged} as follows,
\begin{align}\label{BS-using-F}
\frac{\Gamma(-l-\nu)\Gamma(2l+2)}{\Gamma(-2l)\Gamma(l+1-\nu)}=-(4q)^{2l+1}\frac{F_{2}(\omega)}{F_{1}(\omega)}\underbrace{\left(\frac{1+\frac{B}{A}\frac{F_{2}(-\omega)}{F_{2}(\omega)}}{1+\frac{B}{A}\frac{F_{1}(-\omega)}{F_{1}(\omega)}}\right)}_{\mathcal{G}^{-1}}~,
\end{align}
where the functions $F_{1}(\omega)$ and $F_{2}(\omega)$ have the following expressions,
\begin{align}
F_{1}&=\frac{\Gamma(1-4i\omega)\Gamma(2l+1)}{\Gamma(l+1-2i\omega+\delta)\Gamma(l+1-2i\omega-\delta)}~, 
\label{F-1}
\\
F_{2}&=\frac{\Gamma(1-4i\omega)\Gamma(-2l-1)}{\Gamma(-l-2i\omega+\delta)\Gamma(-l-2i\omega-\delta)}~.
\label{F-2}
\end{align}
In order to derive the bound state frequencies analytically, we have to work in the regime $\omega\sim \omega_{\text{pl}}\ll 1$. So we will keep terms up to linear order in $\omega$, which implies $\delta\simeq 1$. In this case it follows that,
\begin{align}
\frac{F_{1}(-\omega)}{F_{1}(\omega)}\simeq 1+4i\omega \big\{2\Psi(1)-\Psi(l+2)-\Psi(l)\big\}~,
\label{ratio-F-1}
\\
\frac{F_{2}(-\omega)}{F_{2}(\omega)}\simeq -1-4i\omega \big\{2\Psi(1)-\Psi(l+2)-\Psi(l)\big\}~,
\label{ratio-F-2}
\end{align}
where, $\Psi$ denotes the digamma function. Therefore, the function $\mathcal{G}$ becomes,
\begin{align}
\mathcal{G}=\frac{1+\frac{B}{A}\left(1+4i\omega \big\{2\Psi(1)-\Psi(l+2)-\Psi(l)\big\}\right)}{1-\frac{B}{A}\left(1+4i\omega \big\{2\Psi(1)-\Psi(l+2)-\Psi(l)\big\}\right)}~.
\end{align}
Moreover using the reflection formula for the gamma functions, one can show that,
\begin{align} \label{F2/F1}
\frac{F_{2}(\omega)}{F_{1}(\omega)}=\frac{\Gamma(l+2-2i\omega)\Gamma(l+2+2i\omega)\Gamma(l+2i\omega)\Gamma(l-2i\omega)}{\pi (2l)!(2l+1)! \sin(\pi+2\pi l)\, \textrm{cosec}^{2}(l\pi+2i\omega \pi)}~. 
\end{align}
 Note that, we have assumed $l\in \mathbb{C}$, and will take the $l\to \textrm{integer}$ limit at the very end, akin to the analytic continuation method. 
Substituting all of these results in \ref{BS-using-F}, and also incorporating the fact that $\sin(\pi+x)=-\sin x$, we arrive at the following expression,
\begin{equation}\label{simplified-1}
\begin{split}
\frac{\Gamma(-l-\nu)\Gamma(2l+2)}{\Gamma(-2l)\Gamma(l+1-\nu)}=
-(4q)^{2l+1}\frac{\Gamma(l+2-2i\omega)\Gamma(l+2+2i\omega)\Gamma(l+2i\omega)\Gamma(l-2i\omega)}{\pi (2l)!(2l+1)! 
\sin(2 l\pi)~\text{cosec}^{2}(l\pi+2i\omega \pi)}\frac{1}{\mathcal{G}}~.
\end{split}
\end{equation}
In the case of a black hole, the $B=0$ which implies $\mathcal{G}=1$, and \ref{simplified-1} reduces to the corresponding expressions in ~\cite{Rosa2012}. Proceeding similar to~\cite{Rosa2012} and assuming $\nu=\nu_{0}=(l+n+1)$ we obtain,
\begin{align}
\omega\simeq \omega_{\text{pl}}\left(1-\frac{\omega_{\text{pl}}^{2}}{2(n+l+1)^{2}}\right)\equiv \omega_{0}.
\end{align}
\begin{figure}[htbp!]
\includegraphics[width=0.65\linewidth]{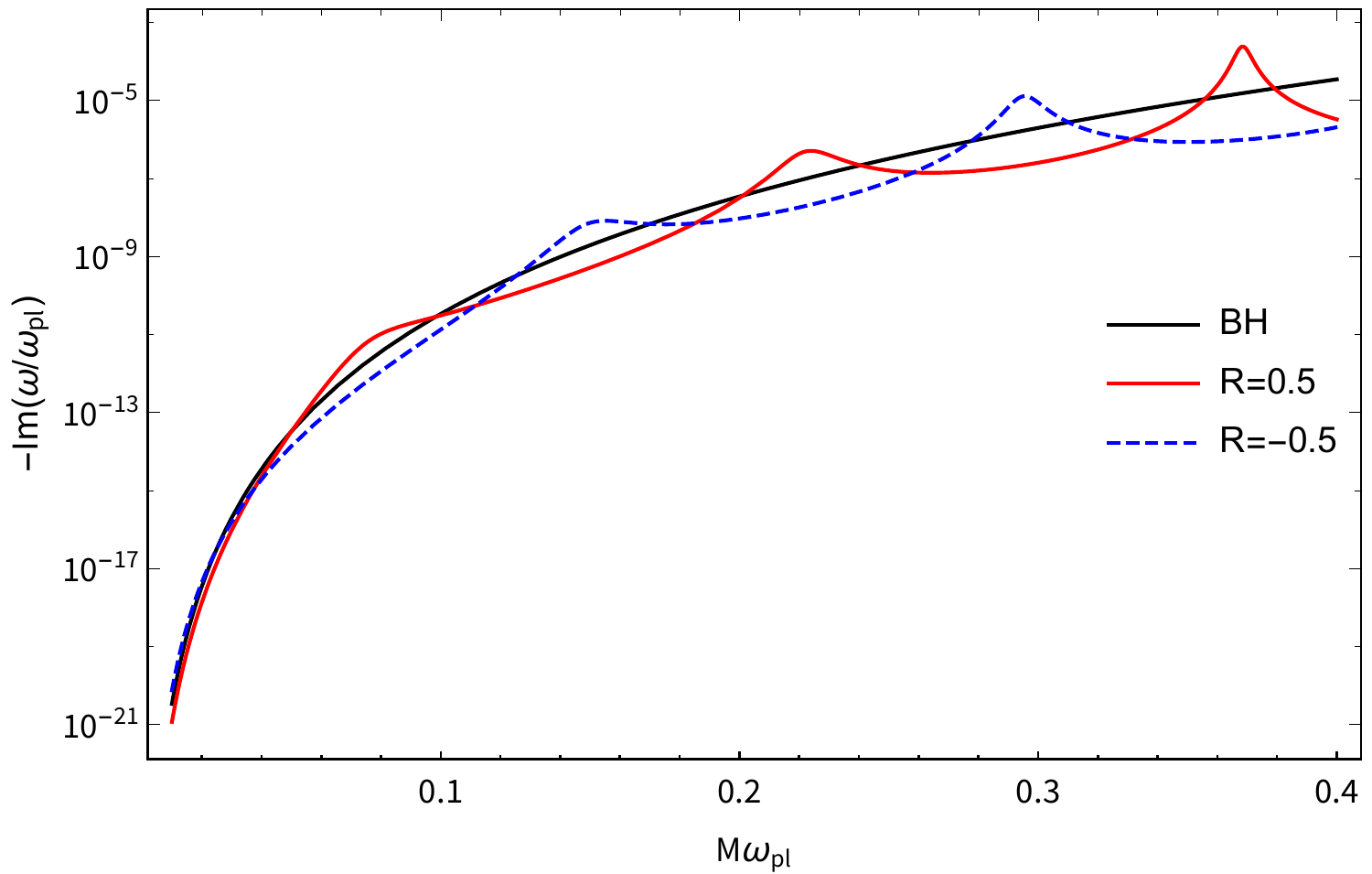}
\caption{The imaginary part of the fundamental plasma-driven axial EM quasi-bound state frequencies has been plotted against the plasma frequency $\omega_{\rm pl}$. We assume homogeneous plasma distribution as well as constant reflectivity R=$\pm 0.5$. The oscillatory behaviour about the BH value is clearly evident.}
\label{fig:analytic}
\end{figure}
This is the real part of the quasi-bound frequency and is independent of the reflectivity of the ECO. This is why the real part of the quasi-bound state frequencies effectively followed the BH trend. To compute the imaginary part of the frequency, we consider $\omega=\omega_{0}+\delta\omega$. Then to the leading order in $\delta\omega$ one can show that, $\delta\nu=-(n+l+1)^{2}(\omega_{\text{pl}})^{-3}\delta\omega$. 
Under this consideration with $q\simeq-\{\omega^{2}_{\text{pl}}/(n+1+l)\}$, one can show that

\begin{align}
\delta\omega \propto 2i\omega_{0} \left[\frac{1-R(\omega_{0})e^{-4i\omega_{0}\ln\epsilon}\left(1+4i\omega_{0} \{2\Psi(1)-\Psi(l+2)-\Psi(l)\}\right)}{1+R(\omega_{0})e^{-4i\omega_{0}\ln\epsilon}\left(1+4i\omega_{0} \{2\Psi(1)-\Psi(l+2)-\Psi(l)\}\right)}\right]~,
\label{eq:analytic-im}
\end{align}
where we have used \ref{def_reflect} and the location of the ECO surface to replace the ratio $(B/A)$ in terms of reflectivity as follows, $(B/A)=R(\omega_0)e^{-4i\omega_0\ln\epsilon}(1-4i\omega_0)$ 
The oscillatory behaviour is due to the $e^{-4i\omega_{0}\ln \epsilon}$ term, with the oscillation scale dependent on $\ln \epsilon$, and the amplitude on $R(\omega_{0})$. Therefore, the features in the numerical spectrum is recovered from our analytical expressions as well. The same can also be seen from the plot of the imaginary part of $\delta \omega$ as well, presented in \ref{fig:analytic}, given the above analytic expression. It is to be emphasized that at large values of $\omega_{\rm pl}$, the numerical and analytical results differ considerably, which has to do with the failure of the assumption $\omega \sim \omega_{\text{pl}}$.
\twocolumngrid
\bibliography{ref}
\bibliographystyle{utphys1}
\end{document}